\documentclass[11pt]{article}
\usepackage{latexsym}
\usepackage{epsfig,empheq}
\usepackage[mathscr]{eucal}
\usepackage{amsfonts}
\usepackage{amscd}
\usepackage{amsmath,amssymb}
\usepackage{amssymb}
\usepackage{colordvi}
\usepackage{graphicx}
\usepackage[footnotesize]{caption}
\usepackage{subcaption}
\usepackage{fancyhdr} 
\usepackage{pdfpages}
\usepackage{slashed}
\usepackage{tabularx}
\usepackage{longtable}
\usepackage{array}
\usepackage{hyperref}
\usepackage[autostyle]{csquotes} 
\usepackage{color}
\usepackage{ulem}

\def\ie{{\it i.e.}}

\def\to{\rightarrow}

\hypersetup{
	colorlinks=true,        % false: boxed links; true: colored links
	linkcolor=blue,          % color of internal links
	citecolor=red,        % color of links to bibliography
	filecolor=magenta,      % color of file links
	urlcolor=cyan           % color of external links
}

\begin{document}

\title{\hfill ~\\[-30mm]
\phantom{h} \hfill\mbox{\small SLAC-PUB-16330} \\[-1.1cm]
\phantom{h} \hfill\mbox{\small TTK--16--03} 
\\[1cm]
\vspace{13mm}   \textbf{Simplified Models for Higgs Physics: Singlet Scalar and Vector-like Quark Phenomenology}}
\date{}
\author{
Matthew J.\ Dolan$^{1}$\footnote{E-mail: \texttt{dolan@unimelb.edu.au}},\;
J.L.\ Hewett$^{2}$\footnote{E-mail: \texttt{hewett@slac.stanford.edu}},\;
M.\ Kr\"amer$^{3}$\footnote{E-mail: \texttt{mkraemer@physik.rwth-aachen.de}}, and 
T.G.\ Rizzo$^{2}$\footnote{E-mail: \texttt{rizzo@slac.stanford.edu}}\\[5mm]
	{\small \it $^1$ARC Centre of Excellence for Particle Physics at the Terascale,}\\{\small \it School of Physics, University of Melbourne, 3010, Australia}\\[1mm]
{\small\it $^2$ SLAC National Accelerator Laboratory, Menlo Park 94025, CA, USA}\\[1mm]
{\small\it 
$^3$Institute for Theoretical Particle Physics and Cosmology,}\\ {\small \it RWTH Aachen University, D-52056 Aachen, Germany}
}

\maketitle

\begin{abstract}
  Simplified models provide a useful tool to conduct the search and exploration of physics beyond the Standard Model in a model-independent fashion. In this work we consider the complementarity of indirect searches for new physics in Higgs couplings and distributions with direct searches for new particles, using a simplified model which includes a new singlet scalar resonance and vector-like fermions that can mix with the SM top-quark. We fit this model to the combined ATLAS and CMS 125 GeV Higgs production and coupling measurements and other precision electroweak constraints, and explore in detail the effects of the new matter content upon Higgs production and kinematics. We highlight some novel features and decay modes of the top partner phenomenology, and discuss prospects for Run II.

\end{abstract}
\thispagestyle{empty}
\vfill
\newpage
\setcounter{page}{1}

\tableofcontents

%%%%%%%%%%%%%%%%%%%%%%%%%%%%%%%%%%%%%%
\section{Introduction}
\label{ch:introduction}
%%%%%%%%%%%%%%%%%%%%%%%%%%%%%%%%%%%%%%%%

Two of the main objectives of the Large Hadron Collider (LHC) Run II are studying the detailed spectroscopy of the Higgs boson and continuing the search for physics Beyond the Standard Model (BSM).  The properties of the recently discovered 125 GeV Higgs boson provide a new tool in the search for new physics and may indirectly reveal its existence, whereas the gain in the center-of-mass energy of the collider in Run II supplies fertile ground to directly discover new particles.    

The majority of searches for new physics at the LHC are now designed and interpreted in terms of Simplified Models~\cite{Alwall:2008ag,Alves:2011wf}. Simplified Models provide a framework for understanding the broad kinematic features of signatures using a small number of parameters, such as masses and couplings of new fields, without depending on specific characteristics of full UV-complete models.  This focus on the relevant weak scale Lagrangian parameters allows for the design of relatively model-independent BSM searches that are broadly applicable to new physics scenarios.  Due to these advantages, the simplified model  approach has recently been extended to searches for dark matter particles at the LHC~\cite{Buchmueller:2013dya,Malik:2014ggro,Buchmueller:2014yoa,Abdallah:2014hon,Abdallah:2015ter}.

The use of simplified models is not restricted to new physics searches, as they can also be of utility in understanding the limits set on new physics from precision Standard Model (SM) measurements such as the properties of the Higgs boson. Another approach often employed in this context is Effective Field Theory (EFT), where constraints are set on the Wilson coefficients of higher dimensional operators constructed out of Standard Model fields. A UV-complete theory can be mapped onto the EFT by integrating out the new heavy states, assuming there is a hierarchy of scales between the SM and new states.  Simplified models differ in allowing the direct exploration of the phenomenology of the BSM states.

The purpose of this paper is to develop the simplified model approach for Higgs physics and investigate the interplay between the various approaches.  We set up a framework to explore BSM theories that affect the Higgs sector and connect measurements of Higgs properties with direct searches for new physics.

To do this we construct a class of simplified models that introduce modifications to the Higgs couplings to SM fields, but which are also amenable to direct searches at the LHC. In particular, we study an extension of the SM that involves a scalar singlet as well as vector-like fermions which mix with the top quark. We argue that such models assist in the exploration of the space of BSM theories that could affect the properties of the Higgs.  Models with this matter content are of interest since singlet scalars can be identified with Higgs portal models~\cite{Patt:2006fw,Englert:2014uua,Robens:2015gla} and vector-like partners of the top-quark appear in numerous BSM scenarios, including composite Higgs models, warped extra dimensions, Little Higgs and extended grand unified theories.  In particular, the presence of a singlet in association with the vector-like quarks is also known to help stabilize the electroweak vacuum~\cite{Xiao:2014kba,Batell:2012zw}.

This scenario leads to calculable changes in the couplings of the SM-like 125 GeV Higgs boson, which we constrain using the combined ATLAS and CMS measurements of the Higgs production cross-sections and branching ratios, as well as other electroweak precision data. 

As well as serving as a framework that illuminates the effects of precision Higgs measurements on extended scalar sectors, this simplified model is useful for exploring vector-like quark phenomenology. The presence of the singlet leads to new top partner decay channels, a fact which has also been noticed in the context of composite Higgs models~\cite{Anandakrishnan:2015yfa,Serra:2015xfa}. There are also a variety of direct searches for the top partners, such as single top-partner production~\cite{Aguilar-Saavedra:2013qpa,Ortiz:2014iza}, which we discuss in detail below. 

Ultimately we find that the EFT and simplified model approaches are complementary. Precision Higgs (and electroweak) measurements set constraints on the parameter space of the simplified model, and in turn inform the collider phenomenology of the new particles\footnote{A similar model has been studied in this context in~\cite{Angelescu:2015kga}.}. 
The constraints on this perturbative simplified model are found to be compatible with the EFT approach~\cite{Brehmer:2015rna}. We show that given the constraints from direct searches that it is unlikely for there to be large effects on Higgs differential distributions, even when the momentum and mass dependence of top-partners running in loops is correctly taken into account.

Recently there has been much interest in a possible diphoton resonance signal at 750~GeV as observed by ATLAS and CMS in the initial few femtobarns data of Run 2~\cite{ATLAS-CONF-2015-081,CMS-PAS-EXO-15-004}. Among other possibilities, a number of analyses have appeared suggesting this may be due to a similar class of models to the one we study here involving singlet scalars and vector-like fermions~\cite{Ellis:2015oso,Falkowski:2015swt,McDermott:2015sck,Benbrik:2015fyz,Gupta:2015zzs,Han:2015dlp,Knapen:2015dap,Zhang:2015uuo,Craig:2015lraxo}. While this further motivates our study, we note that for vector-like top partners the required diphoton signal strength can only be achieved for non-perturbative values of the couplings, unless multiple families of top-partners are present.

The structure of the paper is as follows: In Section~\ref{sec:simp} we introduce and discuss the details of our model, after which we present the indirect constraints from Run I Higgs measurements in Section~\ref{sec:higgs}.  
In Section~\ref{sec:ScalarPheno} we study the singlet scalar phenomenology as well as effects of our model on differential Higgs distributions. We then focus on the vector-like quark phenomenology in Section~\ref{sec:tprime}, discussing current bounds and suggesting new directions to be taken in Run II, after which we conclude.
Some technical details are contained in the Appendix.

%%%%%%%%%%%%%%%%%%%%%%%%%%%%%%%%%%%%%%%%%%%%%5
\section{A Simplified Model}
\label{sec:simp}
%%%%%%%%%%%%%%%%%%%%%%%%%%%%%%%%%%%%%%%%%%%%%%%

In constructing a first simplified model for Higgs physics, we examine a scenario with minimal particle content and interactions that influences the 125 GeV Higgs couplings in a calculable manner. We thus choose to add two ingredients to the SM: ($i$) a scalar singlet $S$, and ($ii$)  a vector-like fermion $F$.  The singlet $S$ acquires a vev, $S=(s+v_S)$, and provides mass for the vector-like fermion, $m_F=y_Fv_S$.  The Higgs and new scalar fields mix via the term $\lambda_{HS}H^\dagger HS^2$, and thus generate new physics effects in both the gauge and fermionic SM Higgs couplings.  Various choices for the quantum numbers of $F$ are possible, and different representations result in specific patterns for the Higgs cross sections and couplings.  These are outlined in Table~\ref{tab:Freps} for the Higgs gauge boson couplings $hV_{\mu\nu}V^{\mu\nu}$ induced by the vector-like fermion $F$. The gauge couplings are defined in terms of the coefficients $\epsilon_V$ as 
\begin{eqnarray}
\gamma\gamma:\,\, \epsilon_\gamma{\alpha\over\pi}{1\over v_H}\Big( {\lambda_{HS}v_H^2\over m_S^2}\Big)\,, 
& & G_aG^a:\,\, \epsilon_g{\alpha_s\over\pi}{1\over v_H}\Big( {\lambda_{HS}v_H^2\over m_S^2}\Big)\,,  \nonumber\\
BB:\,\, \epsilon_B{g'^2\over\pi^2}{1\over v_H}\Big( {\lambda_{HS}v_H^2\over m_S^2}\Big)\,,
& & W_iW^i:\,\, \epsilon_W{g^2\over\pi^2}{1\over v_H}\Big( {\lambda_{HS}v_H^2\over m_S^2}\Big)\,,
\end{eqnarray} 
and the $\epsilon_V$ take on values as determined by the $F$ representation.  Here, $v_H$ is the vaccum expectation value of the SM-like Higgs field $H$.

\begin{table}
\newcommand\T{\rule{0pt}{2.6ex}}       % Top strut
\newcommand\B{\rule[-1.2ex]{0pt}{0pt}} % Bottom strut
		\centering
		\begin{tabular}{|c|c|c|c|c|} \hline
			F & $\epsilon_\gamma$ & $\epsilon_g$ & $\epsilon_B$ & $\epsilon_W$ \\ \hline\hline
			$\left( \begin{array} {c}T' \\ B' \end{array} \right)_{L+R} $ & ${5\over 18}$ &  $- {1\over 6}$ & ${1\over 144}$ & ${1\over 16}$ \T\B \\ 
			$Q_{L+R}$ &  ${1\over 2}Q^2$ &  $-{1\over 12}$ & ${1\over 8}Q^2$ & 0 \T\B \\ 
			$\left( \begin{array}{c} N\\ E \end{array}\right)_{L+R}$ & ${1\over 6}$ &  0 & ${1\over 48}$ & ${1\over 48}$ \T\B \\ 
			$L_{L+R}$   &  ${1\over 16}Q^2$ & 0 & ${1\over 24}$ & 0  \T\B \\ \hline\hline
		\end{tabular}
		\caption{Possible vector-like fermion representations and their influence on the Higgs couplings to gauge fields. $Q^2$
		represents the square of the fermion's electric charge and $\epsilon_V$ is the derived coefficient of the $h V_{\mu\nu}V^{\mu\nu}$ operator.}
		\label{tab:Freps}
	\end{table}

In this paper, for simplicity, we consider the case where $F$ is a color-triplet and SU(2) singlet vector-like fermion
field, $T$, with charge $Q_F = +2/3$. We assume that the vector-like fermion mixes with the SM top-quark only.
This model was chosen such that in the heavy particle limit it
introduces all the (CP-even) dimension-6 operators that affect Higgs physics and are not severely constrained by electroweak precision physics. Such a model has also been considered in \cite{Xiao:2014kba} in the context of stabilizing the electroweak vacuum in the presence of new vector-like fermion representations. We intend on returning to the other possibilities, some of which have been garnering interest recently in the context of the 750~GeV diphoton excess, in future publications.

The most general Lagrangian for our model includes new Yukawa and gauge terms for the SM top-quark and the new vector fermion, as well as an extended  scalar potential:
\begin{equation}\label{eq:lag_gen}
\mathcal{L} \supset \mathcal{L}_{\rm Yukawa} + \mathcal{L}_{\rm gauge} - V(H,S).
\end{equation}
We discuss the various new contributions in Eq.(\ref{eq:lag_gen}) in turn below. 

%%%%%%%%%%%%%%%%%%%%%%%%%%%%%%%%%%%%%%%%%%%%%%%%%%%%%%%
\subsection*{The Yukawa interactions}
%%%%%%%%%%%%%%%%%%%%%%%%%%%%%%%%%%%%%%%%%%%%%%%%%%%%%%%
The most general fermion mass terms for the SM top-quark and the vector-like fermion $T$ are
\begin{eqnarray}
\lefteqn{\hspace*{-5mm}\mathcal{L}_{\rm Yukawa} =  y_T S\overline{T}^{\rm int}_L T^{\rm int}_R + \tilde{\lambda}S\overline{T}^{\rm int}_Lt^{\rm int}_R}\\ \nonumber
&& \qquad + y_t \overline{Q}^{\rm int}_L\widetilde{H} t^{\rm int}_R + y_b \overline{Q}^{\rm int}_L H b_R + \lambda_T \overline{Q}^{\rm int}_L \widetilde{H} T^{\rm int}_R + m_D \overline{T}^{\rm int}_L T^{\rm int}_R,
\end{eqnarray}
where $\widetilde{H} = i\sigma_2 H^*$, $\overline{Q}^{\rm int}_L = (\bar{t}^{\rm int}_L \; \bar{b}^{\rm int}_L)$, and $\psi_{L,R} = \frac12(1\mp\gamma_5)\psi$. We denote the weak interaction eigenstates by $t^{\rm int}$ and $T^{\rm int}$, while $t$ and  $T$ are used for the mass eigenstates.

Through a rotation of the fields $(t^{\rm int}_R, T^{\rm int}_R)$ one can remove the 
term proportional to $ \overline{T}^{\rm int}_L t^{\rm int}_R$. Furthermore, to reduce the number of free parameters, we assume that the mass of the vector fermion is generated from the vev of the singlet field only and set $m_D = 0$. We thus have 
\begin{equation}\label{eq:yuk}
{\mathcal{L}_{\rm Yukawa} = y_T S\overline{T}^{\rm int}_L T^{\rm int}_R + y_t \overline{Q}^{\rm int}_L\widetilde{H} t^{\rm int}_R + y_b \overline{Q}^{\rm int}_L H b_R + \lambda_T \overline{Q}^{\rm int}_L \widetilde{H} T^{\rm int}_R.}
\end{equation}
After spontaneous symmetry breaking, the SM top-quark $t^{\rm int}$ and the vector quark $T^{\rm int}$ mix to form the mass
eigenstates. We  
provide the explicit forms for these mixings in Appendix~\ref{sec:yukawa}.

%%%%%%%%%%%%%%%%%%%%%%%%%%%%%%%%%%%%%%%%%%%%%%%%%%%%%%%
\subsection*{Gauge interactions of the vector-like fermion}
%%%%%%%%%%%%%%%%%%%%%%%%%%%%%%%%%%%%%%%%%%%%%%%%%%%%%%%
The vector-like fermion field $T^{\rm int}$
carries the quantum numbers hypercharge $Y_{T} =  4/3$ and electromagnetic charge $Q_{T} = 2/3$. The terms in the Lagrangian involving the vector-like fermion $T$ and the SM third generation quarks $t, b$ are:
\begin{empheq}
%[box=\fbox]
{align}
 \mathcal{L}_{\rm gauge} &\supset  i\,\bar{t}\slashed{\partial}t  +  i\,\overline{T}\slashed{\partial} T + i\,\bar{b}\slashed{\partial} b \\\nonumber 
&+  g_s \left(\bar{t} \frac{\lambda^k}{2} \gamma^\mu t + \overline{T} \frac{\lambda^k}{2} \gamma^\mu T + \bar{b} \frac{\lambda^k}{2} \gamma^\mu b\right) G_\mu^k\\ \nonumber 
&+  e \left (Q_{t} \bar{t} \gamma^\mu t + Q_{T} \overline{T} \gamma^\mu T + Q_{b} \bar{b} \gamma^\mu b \right)A_\mu\\ \nonumber
&+  \frac{g}{\sqrt{2}}\left (( c_L\bar{t} \gamma^\mu P_L b + s_L\overline{T} \gamma^\mu P_L b)W_\mu^+ +  (c_L \bar{b}\gamma^\mu P_L t + s_L \bar{b}  \gamma^\mu P_L T)W^-_\mu \right )\\ \nonumber
&+  \frac{g}{c_w}\left( \bar{t} \gamma_\mu \left (\frac{c_L^2}{2} P_L - Q_{t}s_w^2\right) t  + \overline{T}\gamma_\mu \left (\frac{s_L^2}{2} P_L - Q_{T}s_w^2\right) T  \right. \\\nonumber 
& + \left. \bar{b} \gamma_\mu \left (-\frac12 P_L - Q_{b}s_w^2\right)b  + \bar{t} \gamma_\mu \frac{s_Lc_L}{2} P_L  T + \overline{T} \gamma_\mu \frac{s_Lc_L}{2} P_L  t \right) Z_\mu, \nonumber
\end{empheq}
with the top-quark vector-like fermion mixing being described by $c_L = \cos\theta_L$ and $s_L = \sin\theta_L$, and the left-handed projector is $P_L = \frac12(1-\gamma_5)$. The cosine and sine of the SM weak mixing angle are denoted by $c_w = \cos\theta_w$ and $s_w = \sin\theta_w$ respectively, and $\lambda_k$ are the Gell-Mann generators of the $SU(3)$ algebra.

%%%%%%%%%%%%%%%%%%%%%%%%%%%%%%%%%%%%%%%%%%%%%%%%%%%%%%%
\subsection*{The scalar potential}\label{sec:scalar}
%%%%%%%%%%%%%%%%%%%%%%%%%%%%%%%%%%%%%%%%%%%%%%%%%%%%%%%
The scalar potential contains the SM Higgs doublet field, $H$, and a real singlet scalar field, $S$:
\begin{equation}\label{eq:potential}
{V(H,S) = -\mu^2 H^\dagger H + \lambda (H^\dagger H)^2 +  \frac{a_1}{2}H^\dagger H \, S + \frac{a_2}{2} H^\dagger H \, S^2\ + b_1 S + \frac{b_2}{2} S^2 + \frac{b_3}{3} S^3 + \frac{b_4}{4} S^4,}
\end{equation}
where 
\begin{equation}\label{eq:fields}
H = 
\begin{pmatrix}
i \phi^+ \\
\frac{1}{\sqrt2} (h + v_H + i\phi^0)
\end{pmatrix} \quad {\rm and} \quad S = (s + v_S).
\end{equation}
Here, we follow the notation of \cite{Chen:2014ask}. 
We quote the most general form of the scalar potential and do not assume a $Z_2$ symmetry for the scalar field $S$ to simplify the potential, as this would also eliminate the $S\overline{T}T$ coupling to the vector-like quark in $\mathcal{L}_{\rm Yukawa}$, see Eq.(\ref{eq:yuk}). It has been suggested  in~\cite{Batell:2012zw} that one could explicitly break the $Z_2$ symmetry by adding a $S\overline{T}T $ term, and that the radiatively generated terms $\propto H^\dagger H \, S$, $S$ and $S^3$ would then have loop-suppressed couplings and could be safely neglected. We shall work out the details of the model using the most general form of the potential, Eq.(\ref{eq:potential}), but shall consider the special case with $a_1 = b_1 = b_3 = 0$ (i.e. we remove the terms which are odd in $S$) for the numerical results presented below. Details of the minimisation of the scalar potential can be found in Appendix~\ref{sec:scalar}.  After mixing, the physical scalar fields are denoted as $h_{1,2}$ with $h_1$ being the SM-like 125 GeV Higgs boson observed at the LHC and $\theta$ representing the mixing angle.

%%%%%%%%%%%%%%%%%%%%%%%%%%%%%%%%%%%%%%%%%%%%%%%%%%%%%%%
\subsection*{Input parameters}
%%%%%%%%%%%%%%%%%%%%%%%%%%%%%%%%%%%%%%%%%%%%%%%%%%%%%%%

In its simplest form, our model has three fixed and five free parameters.  We take the free parameters to be the physics masses of the heavy scalar and top-quark partner, the mixing angles $\theta$ in the scalar, and $\theta_L$ in the fermion sector, respectively, and the vacuum expectation value of the singlet field $v_S$, with the fixed parameters being the mass and vev of the SM-like Higgs as well as the top-quark mass:
\begin{eqnarray}\label{eq:inputparams}
&m_{h_2}, \, \theta, \, v_S, \, m_T \; {\rm and}\; \theta_L ; \\ \nonumber
&m_{h_1} = 125.0~{\rm GeV}, v_H = 246~{\rm GeV},  m_t = 173.2~{\rm GeV}\,.
\end{eqnarray}

We neglect the effects on the Higgs self-couplings as the LHC will only have modest sensitivity to these with a very large amount of integrated luminosity~\cite{Dolan:2012rv,Goertz:2013kp,Barr:2013tda,Dolan:2015zja}, although some evidence of these could be accessible through resonant heavy Higgs production~\cite{Dolan:2012ac,Chen:2014ask}.  
The input parameters Eq.(\ref{eq:inputparams}) are related to the Lagrangian parameters of $\mathcal{L} \supset \mathcal{L}_{\rm Yukawa} + \mathcal{L}_{\rm gauge} - V(H,S)$ as detailed in Appendix~\ref{sec:input}.
The free parameters of our simplified model are constrained by perturbative unitarity and electroweak precision data, as well as the Higgs cross sections and branching ratios as we will see below.

\section{Higgs Couplings and Constraints}
\label{sec:higgs}

We are primarily interested in the couplings of the light scalar $h_1$ (the 125 GeV Higgs boson) to the Standard Model fermions of the third generation ($t,b,\tau$) and the gauge bosons ($V=W^\pm,Z,\gamma,g$), parametrized in terms of the scale factors $\kappa \equiv g/g^{\rm SM}$:
\begin{eqnarray}
\mathcal{L}_{\rm Higgs}  &  = & \kappa_W\, g_{hWW}^{\rm SM}\, h_1 W^{+\mu}W^-_\mu + \kappa_Z\, g_{h ZZ}^{\rm SM}\, h_1 Z^\mu Z_\mu\\ \nonumber
& - &   \kappa_t\, g_{htt}^{\rm SM}\, h_1 \bar{t}t   - \kappa_b\, g_{hbb}^{\rm SM} \, h_1 \bar{b}b  - \kappa_\tau \, g_{h\tau\tau}^{\rm SM} \, h_1 \bar{\tau}\tau \\ \nonumber
& + &   \kappa_g\, g_{hgg}^{\rm SM} \, h_1 G^{\mu\nu} G_{\mu\nu}   +  \kappa_\gamma \, g_{h\gamma\gamma}^{\rm SM}\,  h_1 A^{\mu\nu} A_{\mu\nu}  . \nonumber 
\end{eqnarray}
The Standard Model tree-level couplings are
\begin{equation}
g^{\rm SM}_{hWW} =  \frac{2m_W^2}{v_H},\quad
g^{\rm SM}_{hZZ}  =  \frac{m_Z^2}{v_H},\quad {\rm and} \quad
g^{\rm SM}_{hff} =  \frac{m_f}{v_H}.
\end{equation}
The couplings to the photon and the gluon are loop-induced. At the one-loop level, they are given by 
\begin{eqnarray*}
g_{hgg}^{\rm SM} & = & \frac{g_s^2}{4 \pi^2} \sum_f \frac{g^{\rm SM}_{hff}}{m_f}A_{1/2}(\tau_f),\\
g_{h\gamma\gamma}^{\rm SM} & = & \frac{e^2}{4 \pi^2}\left(\frac{g^{\rm SM}_{hWW}}{m_W^2}A_1(\tau_W) +  \sum_f 2 N^f_C Q_f^2 \frac{g^{\rm SM}_{hff}}{m_f}A_{1/2}(\tau_f)\right),
\end{eqnarray*}
with $\tau = 4 m_f^2/m_{h_1}^2$, and the loop functions
\begin{eqnarray*}
A_{1/2}(\tau) & = & 2 \tau (1+(1-\tau)f(\tau)),\\
A_{1}(\tau) & = & -2 -3 \tau(1+(2-\tau)f(\tau)), 
\end{eqnarray*}
where 
\begin{equation*}
f(x) 
= \left\{
\begin{array}{cc}
\arcsin^2(1/\sqrt{x}) & {\rm for}\; x\ge 1\\
-\frac14\left(\ln \frac{1+\sqrt{1-x}}{1-\sqrt{1-x}} -i\pi  \right)^2 & {\rm for}\; x < 1\;.
\end{array}\right.
\end{equation*}

In our model, the tree-level couplings to $V=W^\pm,Z$ and to the fermions other than the top-quark, are only modified due to the mixing of the Higgs fields $h$ and $s$, $h_1 = c_\theta\, h - s_\theta\, s$, where $c_\theta = \cos\theta$ and $s_\theta = \sin\theta$, with $\theta$ as given by Eq.(\ref{eq:theta}) in the Appendix, as the singlet field $s$ does not couple directly to the SM fields. Thus we have 
\begin{equation} 
\kappa_W = \kappa_Z = \kappa_b = \kappa_\tau = \cos\theta\,.
\end{equation}

The couplings to the top-quark are modified through both the mixing of the scalar fields and the mixing in the top-quark sector. From the Yukawa-Lagrangian given in the Appendix, Eq.(\ref{eq:L_yuk}), we find 
\begin{eqnarray}
\lefteqn{\mathcal{L}_{\rm Yukawa}  \supset  
(\bar{t}_L \overline{T}_L)\, \mathcal{U}_L (\mathcal{H} + \mathcal{S})\,
\mathcal{U}_R^\dagger
\begin{pmatrix}
t_R \\ T_R 
\end{pmatrix}}\\ \nonumber
&& \hspace*{-8mm} = 
(\bar{t}_L \overline{T}_L)\left(
 \frac{h-i\phi^0}{v_H}
\begin{pmatrix}
m_t c_L^2 & m_T s_L c_L \\ 
m_t s_L c_L & m_T s_L^2 
\end{pmatrix} + 
\frac{s}{v_S}
\begin{pmatrix}
m_t s_L^2 & -m_T s_L c_L \\ 
-m_t s_L c_L & m_T c_L^2 
\end{pmatrix}\right)
\begin{pmatrix}
t_R \\ T_R 
\end{pmatrix}\\ \nonumber
&= & \frac{m_t}{v_Hv_S}\left(c_L^2v_S\,(h-i\phi^0) + s_L^2v_H\,s\right)\bar{t}_L t_R\\ \nonumber
&+ & \frac{m_T}{v_Hv_S}\left(s_L^2v_S\,(h-i\phi^0) + c_L^2v_H\,s\right)\overline{T}_L T_R\\ \nonumber 
& + & \frac{m_T}{v_Hv_S}s_Lc_L\left(v_S\,(h-i\phi^0) - v_H\,s\right)\bar{t}_L T_R\\ \nonumber
& + & \frac{m_t}{v_Hv_S}s_Lc_L\left(v_S\,(h-i\phi^0) - v_H\,s\right)\overline{T}_L t_R. \nonumber 
\end{eqnarray}
For the couplings of the light scalar, $h_1$, with the light and heavy top-quarks, $t, T$, we thus find 
\begin{eqnarray*} 
\lefteqn{\hspace*{-10mm}\mathcal{L}_{\rm Yukawa} \supset 
\frac{m_t}{v_Hv_S}\left(c_L^2c_\theta v_S - s_L^2s_\theta v_H\right)\bar{t}_L t_R h_1 + \frac{m_T}{v_Hv_S}\left(s_L^2c_\theta v_S\ - c_L^2s_\theta v_H\right)\overline{T}_L T_R h_1}\\ \nonumber
&& \hspace*{-2mm}+ \frac{m_T}{v_Hv_S}s_Lc_L\left(c_\theta v_S + s_\theta v_H\right)\bar{t}_L T_R h_1+ \frac{m_t}{v_Hv_S}s_Lc_L\left(c_\theta v_S + s_\theta v_H\right)\overline{T}_L t_R h_1, \nonumber
\end{eqnarray*}
and therefore 
\begin{equation}
\kappa_t = c_L^2c_\theta - s_L^2s_\theta\frac{v_H}{v_S}.
\end{equation}
It is also straightforward to calculate the couplings of the heavier resonance $h_2$ using the above.
The loop-induced couplings to the gluon and the photon differ from those of the Standard Model because of the modified Higgs $W$ and top couplings as well as the additional contribution of the heavy-top loop:
\begin{eqnarray}
g_{h_1gg} & = & \frac{g_s^2}{4 \pi^2}\left(\sum_f \frac{g_{h_1ff}}{m_f}A_{1/2}(\tau_f) + \frac{g_{h_1TT}}{m_T}A_{1/2}(\tau_T)\right),\,\\ \nonumber 
g_{h_1\gamma\gamma} & = & \frac{e^2}{4 \pi^2}\left(\frac{g_{h_1WW}}{m_W^2}A_1(\tau_W) +  \sum_f 2 N^f_C Q_f^2 \frac{g_{h_1ff}}{m_f}A_{1/2}(\tau_f)+  \frac83 \frac{g_{h_1TT}}{m_T}A_{1/2}(\tau_T)\right),
\end{eqnarray}
with $g_{h_1WW} = c_\theta\, g_{hWW}^{\rm SM}$, 
$g_{h_1ff} = c_\theta\, g_{hff}^{\rm SM}$ if $f \neq t$, $g_{h_1tt} = (c_L^2c_\theta - s_L^2s_\theta\frac{v_H}{v_S})\,g_{htt}^{\rm SM}$ and $g_{h_1TT} = (s_L^2c_\theta - c_L^2s_\theta\frac{v_H}{v_S})\,\frac{m_t}{m_T}\,g_{htt}^{\rm SM}$.

We can provide an approximation to the Higgs-gluon coupling by neglecting contributions from fermions other than the top-quark, and considering the limit $m_{h_1} \ll m_t, m_T$, or $\tau_{t,T} \gg 1$. With  $A_{1/2}(\tau) = \frac43 (1 + \frac{7}{30} \tau^{-1} + \mathcal{O}(\tau^{-2}))$ we have
\begin{eqnarray}
g_{h_1gg} & \approx & \frac{g_s^2}{4 \pi^2}\left(\frac{g_{h_1tt}}{m_t}A_{1/2}(\tau_t) + \frac{g_{h_1TT}}{m_T}A_{1/2}(\tau_T)\right)\\ \nonumber
 & \approx & \frac{g_s^2}{4 \pi^2}\frac43 \left(\frac{g_{h_1tt}}{m_t} + \frac{g_{h_1TT}}{m_T}\right)\\ \nonumber
 & = & \frac{g_s^2}{4 \pi^2}\frac43 \frac{g_{htt}^{\rm SM}}{m_t}\left(c_L^2c_\theta - s_L^2s_\theta\frac{v_H}{v_S} + s_L^2c_\theta - c_L^2s_\theta\frac{v_H}{v_S}\right)\\ \nonumber
  & \approx & g_{hgg}^{\rm SM} \left(c_\theta - s_\theta\frac{v_H}{v_S} \right)\,, \\ \nonumber
\end{eqnarray}
and therefore 
\begin{equation}
\kappa_g \approx c_\theta - s_\theta\frac{v_H}{v_S}.
\end{equation}

Since the Higgs couplings to the $W$ and $Z$ bosons are the most accurately measured to date, they will provide the strongest constraints on the mixing angle $\theta$ (especially since the photonic couplings are affected by the presence of the top partners $T$).
We can also  derive a limit on $\theta_L$, and a weak bound on $m_T$, from $\kappa_g$ and 
$\kappa_\gamma$, although the dependence on these parameters in these loop-induced modes is mild. The 5th parameter of the model, the mass of the heavy scalar $m_{h_2}$, does not enter the 125 GeV Higgs couplings to Standard Model fermions and gauge bosons and we will find it to be less constrained by Higgs searches.

In a similar fashion, we can also write simple approximate expressions for the photonic coupling 
ratio $\kappa_\gamma$ as
\begin{equation}
	\kappa_\gamma \approx  \frac{c_\theta A_1(\tau_W)+{{16}\over {9}}\kappa_g}{A_1(\tau_W)+{{16}\over {9}}} \,,
\end{equation}
where $\kappa_g$ and the loop function $A_1$ are given above. We also can obtain a simple expression 
for the triple-Higgs coupling, $\kappa_{3h_1}$, which is given by 
\begin{equation}
	\kappa_{3h_1} \approx c_\theta^3-{{v_H}\over {v_S}} s_\theta^3\,.
\end{equation}
As current projections are that this quantity will only be constrained at the LHC at the 30\% level, this is will not provide useful constraints on the model. 

\begin{figure}[tb]
	\centerline{\includegraphics[width=0.60\textwidth,angle=0]{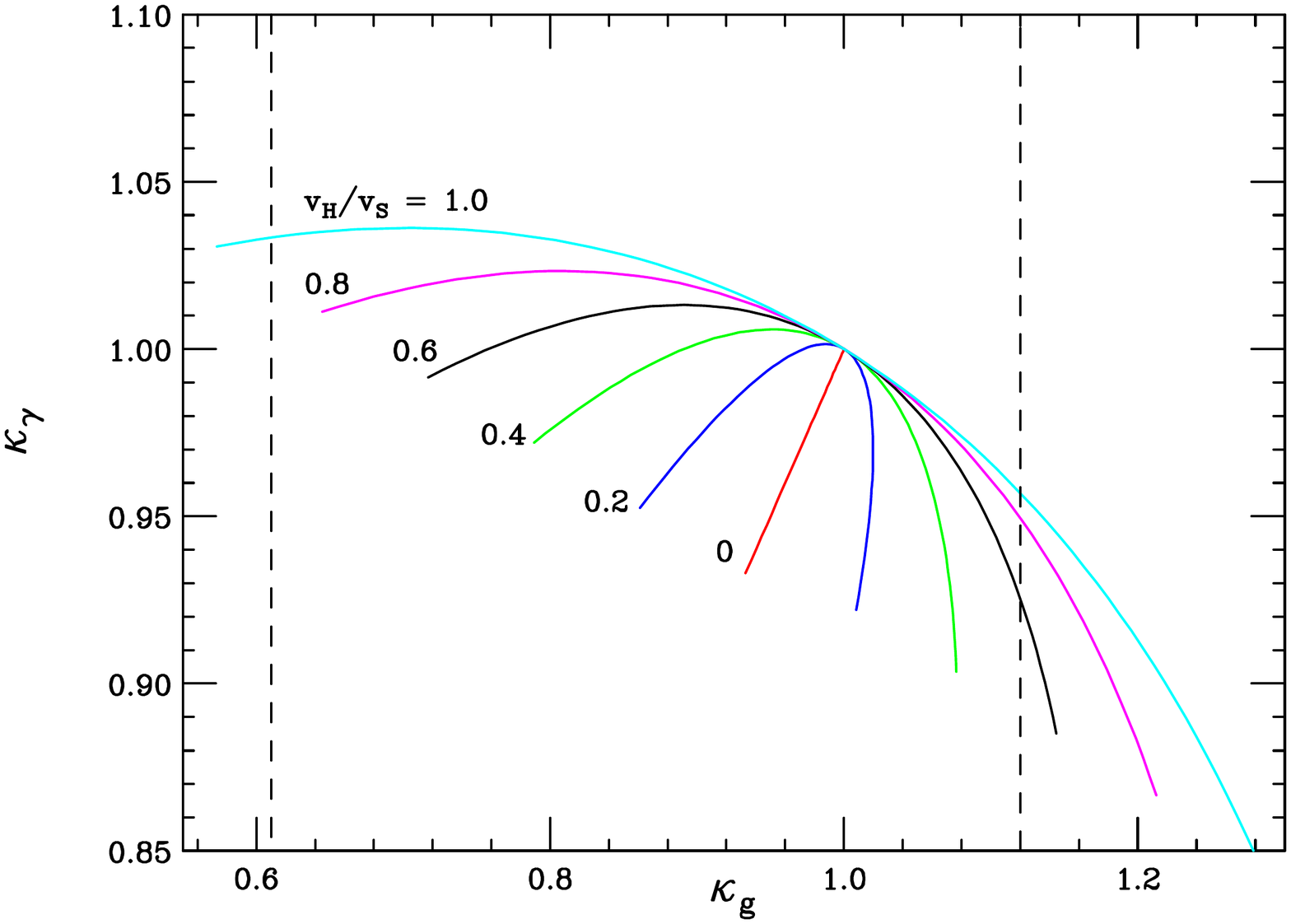}
		\hspace*{-2.0cm}
		\includegraphics[width=0.62\textwidth,angle=0]{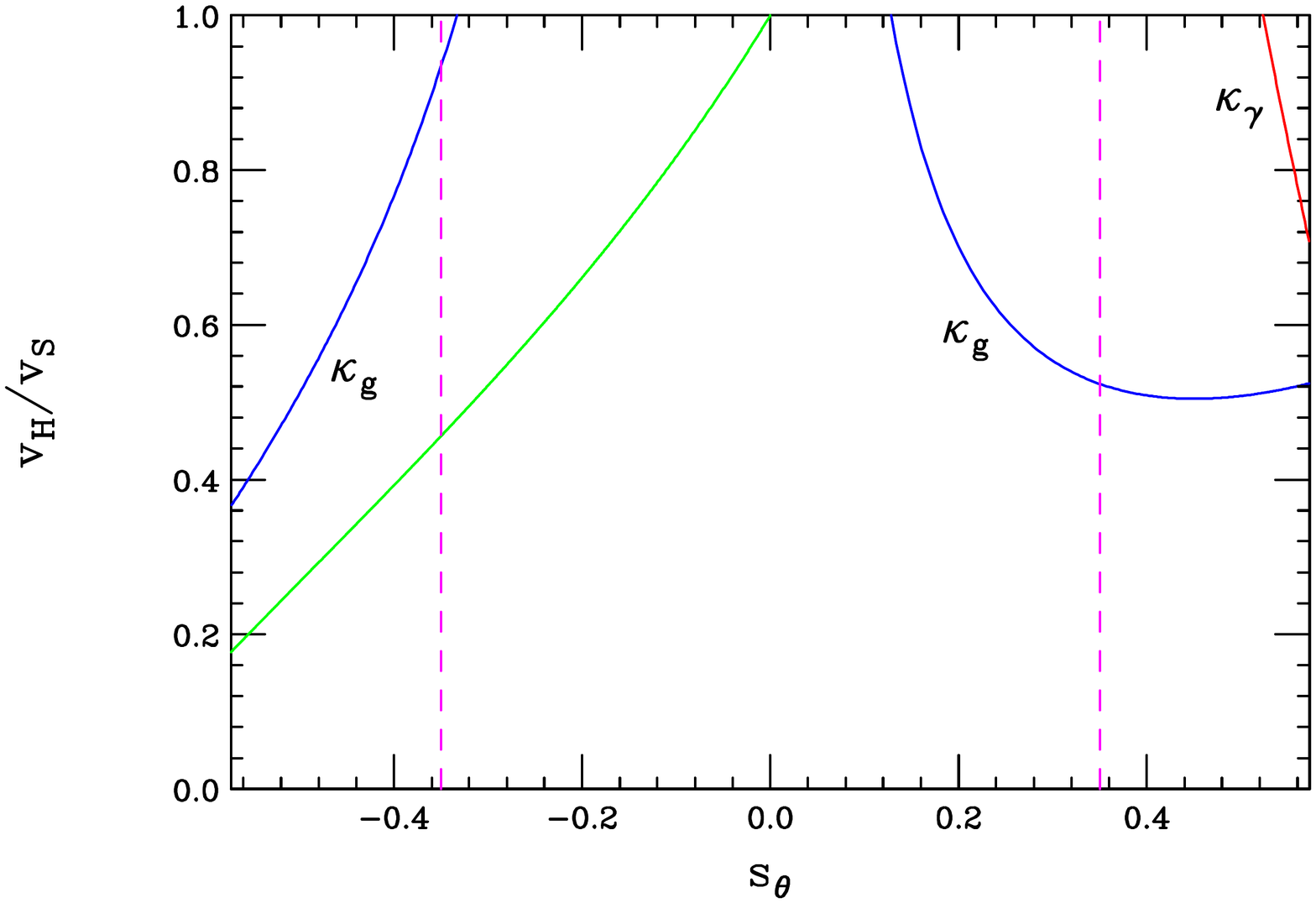}}
	\vspace*{-1.0cm}
	\caption{(Left) The region of the $\kappa_g$-$\kappa_\gamma$ plane that is allowed in the simplified 
		model discussed in the text. From outside going inwards (cyan to red) the curves correspond to values of 
		$r\equiv v_H/v_S$ 
		ranging from 1 to zero in steps of 0.2. The value of $s_\theta$ varies along each curve over the range 
		-0.35 to 0.35, roughly the range allowed by our fit below,  with $s_\theta=0$ being located at 
		$\kappa_g=\kappa_\gamma=1$. The vertical dashed lines represent the $95\%$ CL bounds 
		from the combined ATLAS/CMS 
		Higgs coupling analysis.  (Right) Constraints in the $s_\theta-v_H/v_S$ parameter plane from 
		determinations of $\kappa_g$ 
		(blue) and $\kappa_\gamma$ (red) from the ATLAS/CMS Higgs coupling analysis. The allowed region lies below 
		the curves. The values of $s_\theta=\pm 0.35$, their approximate limit from our fit below, are also shown as the
		dashed lines. As discussed in the text, the region above the green curve is where $|g_{h_2Tt}/g_{h_1Tt}|>1$.}
	\label{example}
\end{figure}

Before discussing our global fit to the 5 parameters present in this simplified model, it is 
useful to understand the effect mixing has on 
the SM-like Higgs properties and what can be learned in the future from more precise measurements. Clearly, as we saw above, 
since $\kappa_{W,Z,b,\tau}=c_\theta \leq 1$, determinations of these quantities in excess of unity are 
outside of the parameter space allowed by our simplified model. If any such values were obtained the model would 
be excluded. For the loop-induced couplings $\kappa_{g,\gamma}$ the situation is less straightforward since both of these 
quantities depend upon the parameters $s_\theta$ and $r\equiv v_H/v_S$ in different, yet correlated, ways. 
In the left panel of Fig.~\ref{example}, we show the region in the $\kappa_g-\kappa_\gamma$ plane that is accessible 
in our simplified model assuming that $|s_\theta| \leq 0.35$, roughly the range allowed by our global fit to be discussed 
below. Each curve corresponds to a fixed value of $0 \leq r \leq 1$ while the value of $s_\theta$ is allowed to vary 
along each of the curves; it is clear that the full allowed region is contained to an ellipse in this plane. 
Here we see that values of $\kappa_g >1$ are associated with values of 
$\kappa_{\gamma} <1$, and vice versa, and that
both parameters cannot simultaneously exceed unity. We also observe that 
values of $\kappa_{\gamma} \gtrsim 1.04$ are not allowed in this simplified model. Again, a measurement of 
either or both of these two quantities outside the locus of points shown here would exclude this model.

The right hand panel of Fig.~\ref{example} shows the impact of the ATLAS/CMS determinations~\cite{ATLAS-CONF-2015-044} of $\kappa_{g,\gamma}$ 
in the $s_\theta-v_H/v_S$ parameter plane. The allowed region from the $\kappa_{g,(\gamma)}$ constraint 
lies below the blue (red) curve while the vertical dashed lines show the approximate range of $|s_\theta| \leq 0.35$ 
allowed by our global fit described below. 
Of particular interest, for discussion later below, is the region above the green curve where we 
find that the ratio of the $h_2Tt$ and $h_1Tt$ couplings satisfies
\begin{equation}
	{{g_{h_2Tt}^2}\over {g_{h_1Tt}^2}} = \Big ({{s_\theta-rc_\theta}\over {c_\theta+rs_\theta}}\Big)^2 >1\,,
\end{equation}
so that the decay rate for $T\to h_2t$ can naively be potentially comparable to that for $T\to h_1t$.
	
We now perform a global fit of this simplified model to the Higgs and electroweak precision data.	
We use the \texttt{HiggsBounds4.2.0} and \texttt{HiggsSignals1.4.0} programs~\cite{Bechtle:2008jh,Bechtle:2013xfa,Bechtle:2013wla} and the MultiNest algorithm~\cite{Feroz:2008xx} to fit the
ATLAS and CMS Higgs data presented in Table 10 of~\cite{ATLAS-CONF-2015-044} to our model. As the HiggsBounds program returns whether a parameter point is ``allowed'' or ``not allowed'', we  take $\chi^2_{\rm{allowed}}=0$ and $\chi^2_{\rm{not\,\, allowed}}=10^4$. We set the MultiNest parameters to $n_{live}=20,000$ and $tol=10^{-3}$ in order to adequately explore the likelihood and parameter space of the model. We also include the Peskin-Takeuchi electroweak precision 
parameters~{\cite{Peskin:1991sw}}, denoted by $S,T$, in the fit, along with the effects of the correlation 
matrix~\cite{Xiao:2014kba,He:2001fz}.  To be specific, we employ the new Higgs singlet 
contributions to both $S$ and $T$ as well as those from the isosinglet quark contributions to $T$ as given in the first of 
these references. The isosinglet quark contributions are taken from the second reference with a correction made to their 
Eq.(90) where $2y_1y_2 \to 22y_1y_2$. Our fit makes use of the values and correlations as presented by 
Cuichini~{\cite {Cuichini}}. Note that our fit does not include the effects of direct searches for the $T$-quark from the LHC so that we may make probabilistic statements about the novel $T\to h_2 t$ decays which appear in our model, and their effects on standard LHC limits. We focus in detail on the $T$-quark phenomenology below in Section~\ref{sec:tprime}.

\begin{figure}[t!]
		\vspace{0.3cm}
%		\parbox{0.95\textwidth}
			\centerline{\includegraphics[width=0.4\textwidth]{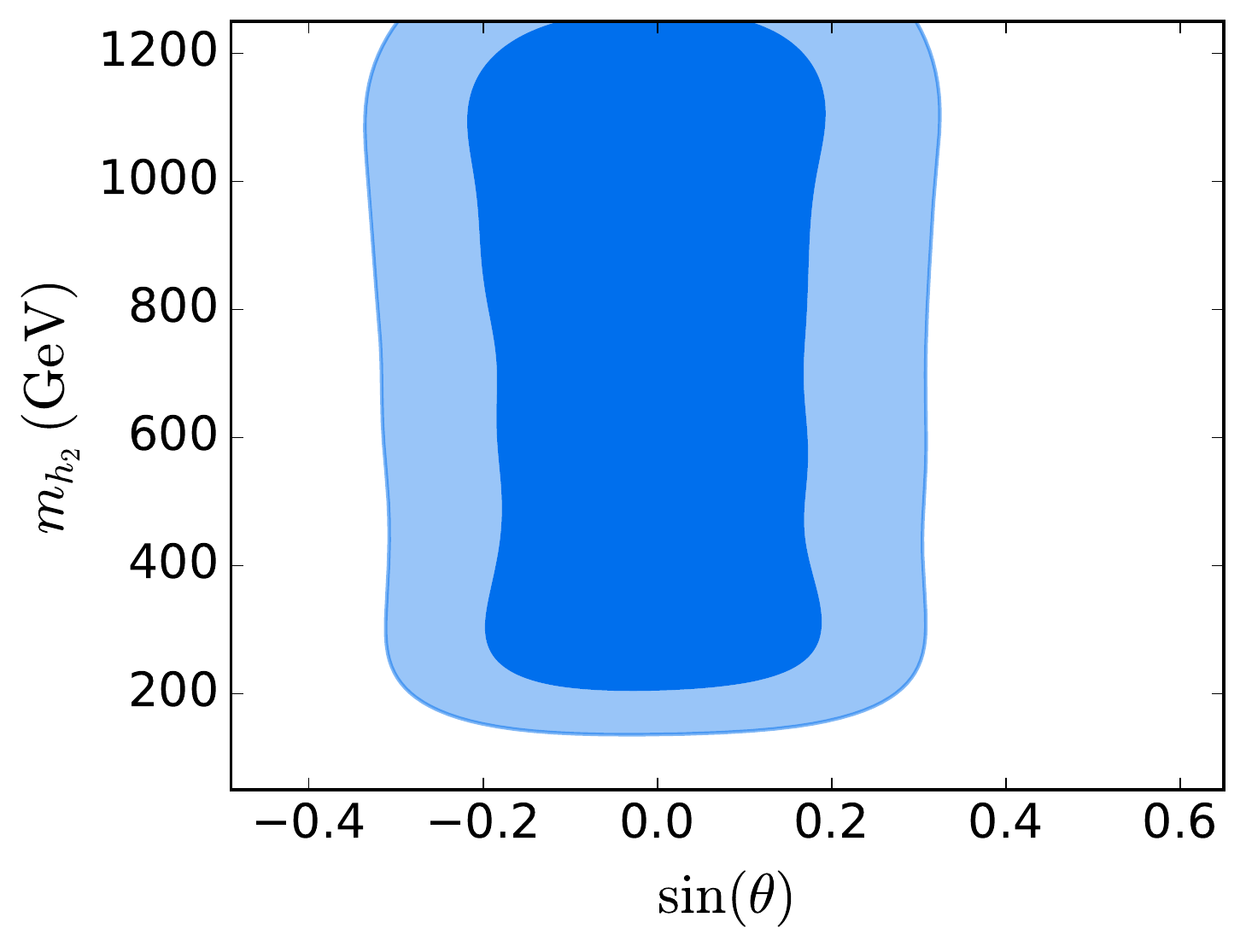}
			\includegraphics[width=0.4\textwidth]{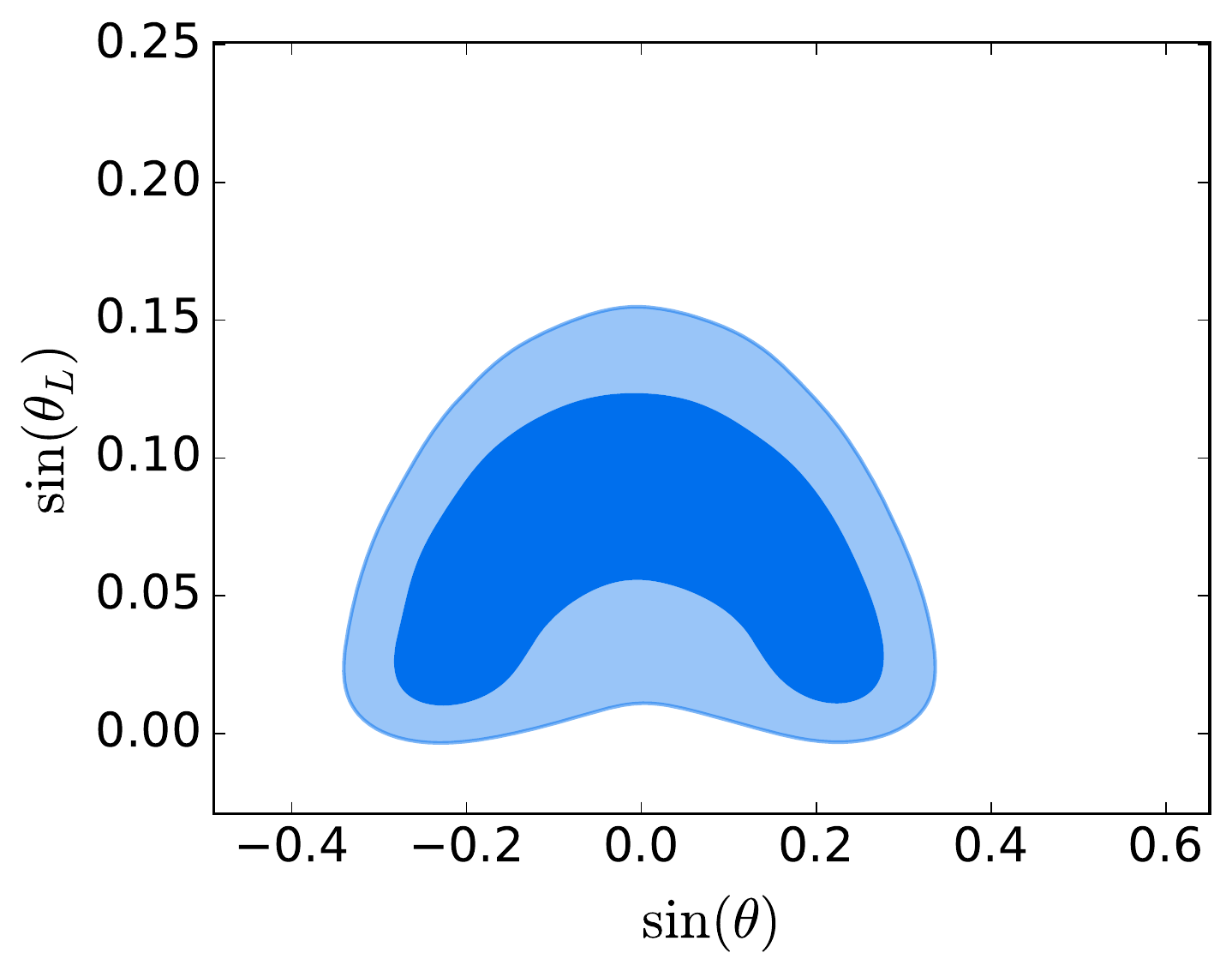}}
		\centerline{\includegraphics[width=0.4\textwidth]{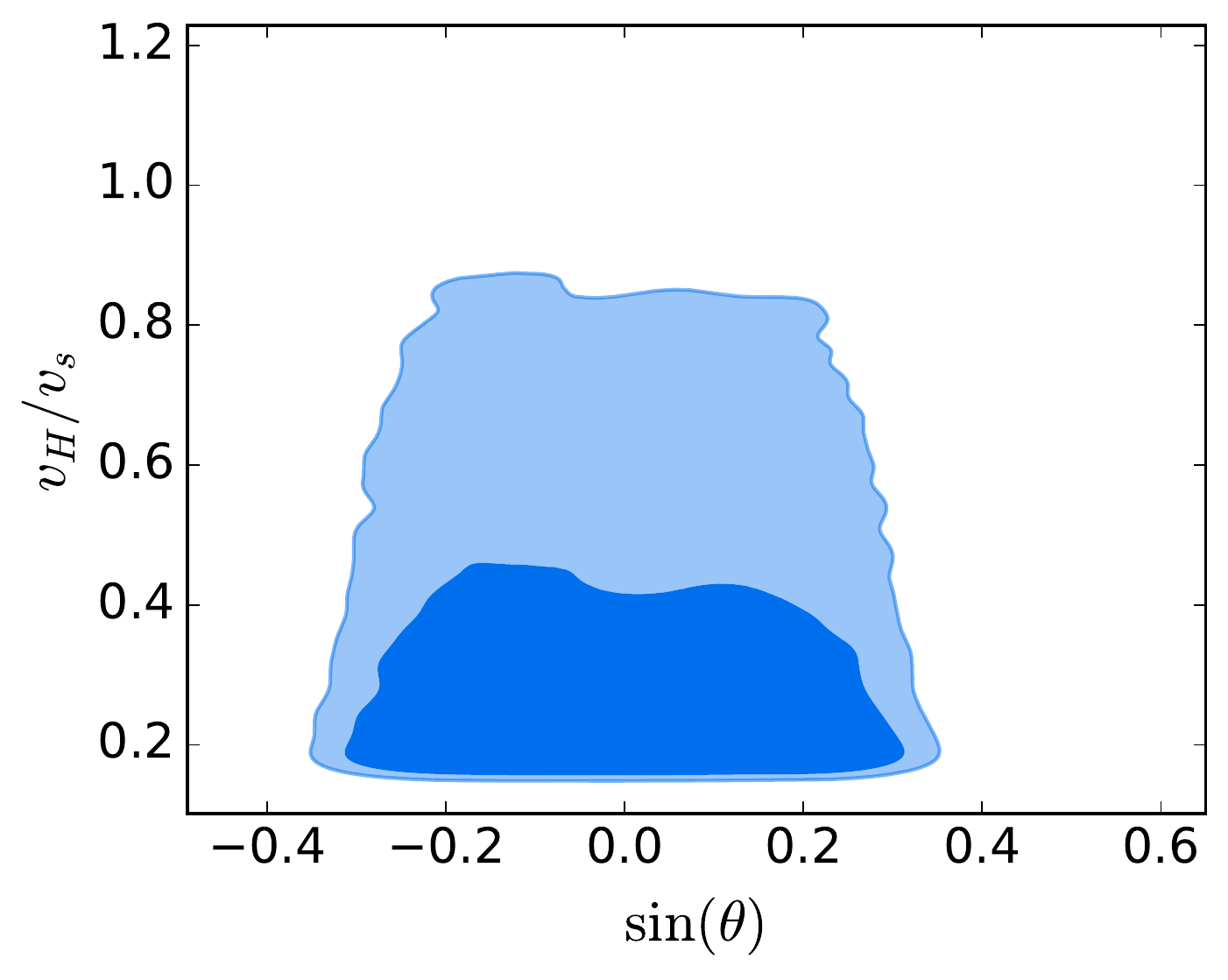}
			\includegraphics[width=0.4\textwidth]{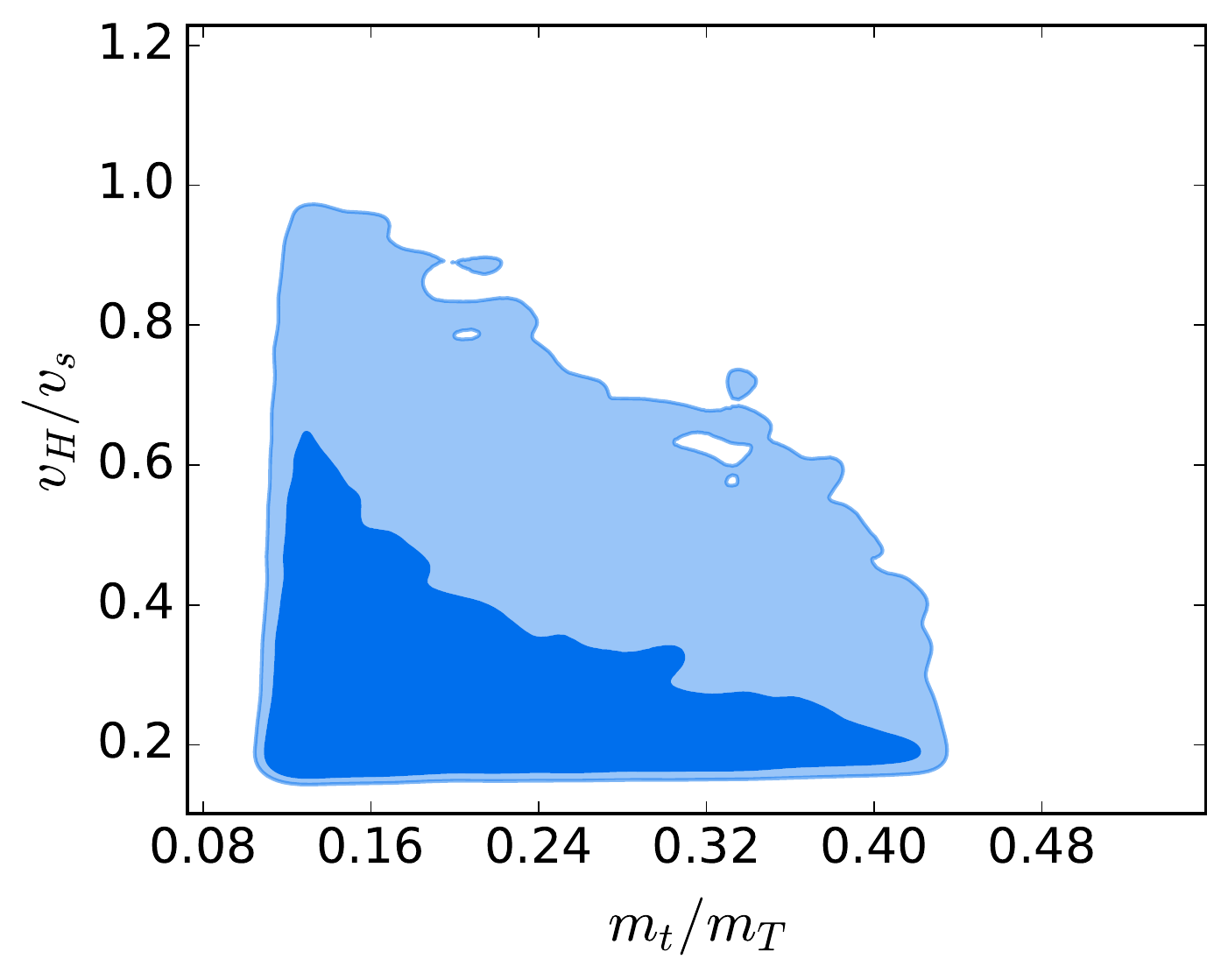}}
			\caption{ 68 and 95\% Bayesian confidence limit contours (light and dark blue respectively) obtained from the fit to Higgs and precision electroweak data in the (clockwise from top-left) $\sin\theta$-$m_{h2}$ , $\sin\theta$-$\sin\theta_L$,  $\sin\theta$-$v_H/v_S$ and $m_t/m_T$-$v_H/v_S$ parameter planes.}
			\label{fig:higgsfit}
		
	\end{figure}

The results of the fit are shown in Fig.~\ref{fig:higgsfit} and Tab.~\ref{tab:higgsfit}, which also shows the ranges of the parameters we scan over as well as the best-fit values for the parameters. Although we scan over the parameters in Tab.~\ref{tab:higgsfit}, we present our results in terms of $\sin\theta$, $\sin\theta_L$, $m_t/m_T$ and $v_H/v_S$ since these are the quantities which enter the physical expressions discussed above.
We see that the distribution in $m_{h_2}$ is relatively flat for $m_{h_2}$ above $\sim 200$\;GeV, since once $\sin\theta$ is small the $h_2$ production cross-sections are suppressed, evading the LHC searches and limits.
In turn, $\sin\theta$ is forced to be close to 0 ({\it i.e.},~towards the SM limit) due to the good agreement between the experimentally measured Higgs properties and those predicted in the SM. The $\sin\theta$ distribution is slightly asymmetric, and the best-fit point has $\sin\theta<0$, with an overall very slight preference for $\sin\theta<0$ compared to $\sin\theta>0$.
Both the ratios $v_H/v_S$ and $m_t/m_T$ are already driven towards the decoupling limit by current measurements of the Higgs properties (as well as the $S$ and $T$ parameters), with 68\% confidence intervals restricting $v_H/v_S\leq 0.85$ and $m_t/m_T\leq 0.42$.

	\begin{table}
		\centering
		\begin{tabular}{|c|c|c|} \hline
			Parameter & Range & Best fit value \\ \hline\hline
			$\theta$ & $\left( -\pi/2, \pi/2 \right)$ &  -0.16\\ \hline
			$v_S$~(GeV) &  $\left( 250.0, 1500.0 \right)$ &  250~GeV\\ \hline
			$m_T$~(GeV) & $\left(  250.0,1500.0 \right)$ &  400~GeV\\ \hline
			$\theta_L$ &  $\left(0,\pi/2 \right)$  & $0.11$ \\ \hline
			$m_{h2}$~(GeV)   &  $\left( 150.0, 1250.0 \right)$ & 1230~GeV \\ \hline
		\end{tabular}
		\caption{The parameter scan ranges used in the fit to the Higgs data as described in the text, and the best fit values obtained from the fit.}
		\label{tab:higgsfit}
	\end{table}

%%%%%%%%%%%%%%%%%%%%%%%%%%%%%%%%%%%%%%%%%%%%
\section{Collider Phenomenology of the Scalar Sector}
\label{sec:ScalarPheno}
%%%%%%%%%%%%%%%%%%%%%%%%%%%%%%%%%%%%%%%%%%%%%%

In this section, we examine the phenomenology of the $h_{1,2}$ scalar bosons at the LHC.  We first explore effects of $h$-$S$ 
mixing on $h_1$ production rates to determine the degree of difference from Standard Model expectations and then study
$h_2$ production and decay. 

\begin{figure}[tb]
\centerline{\includegraphics[bb = 150 535 500 800, clip, width=0.475\textwidth]{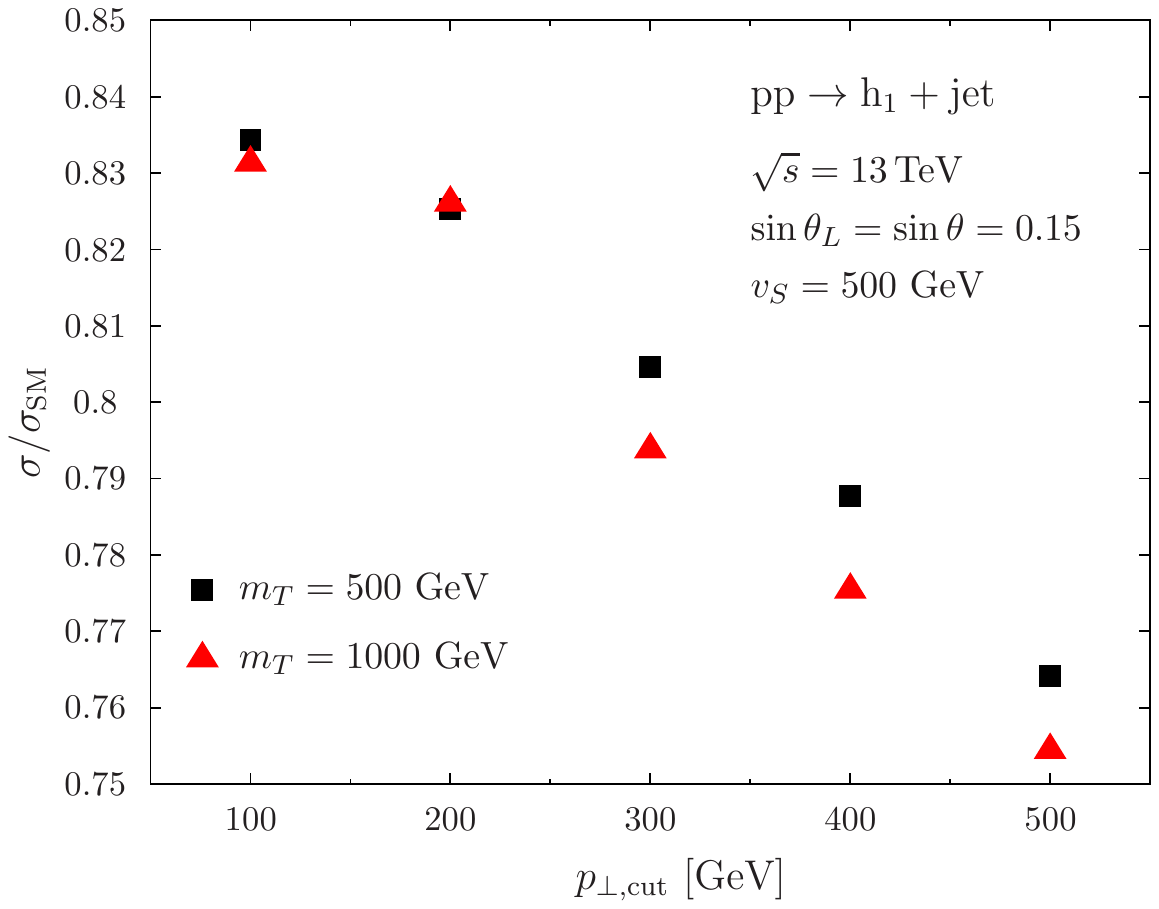}
\hspace*{-0.0cm}
\includegraphics[bb = 150 535 500 800, clip, width=0.475\textwidth]{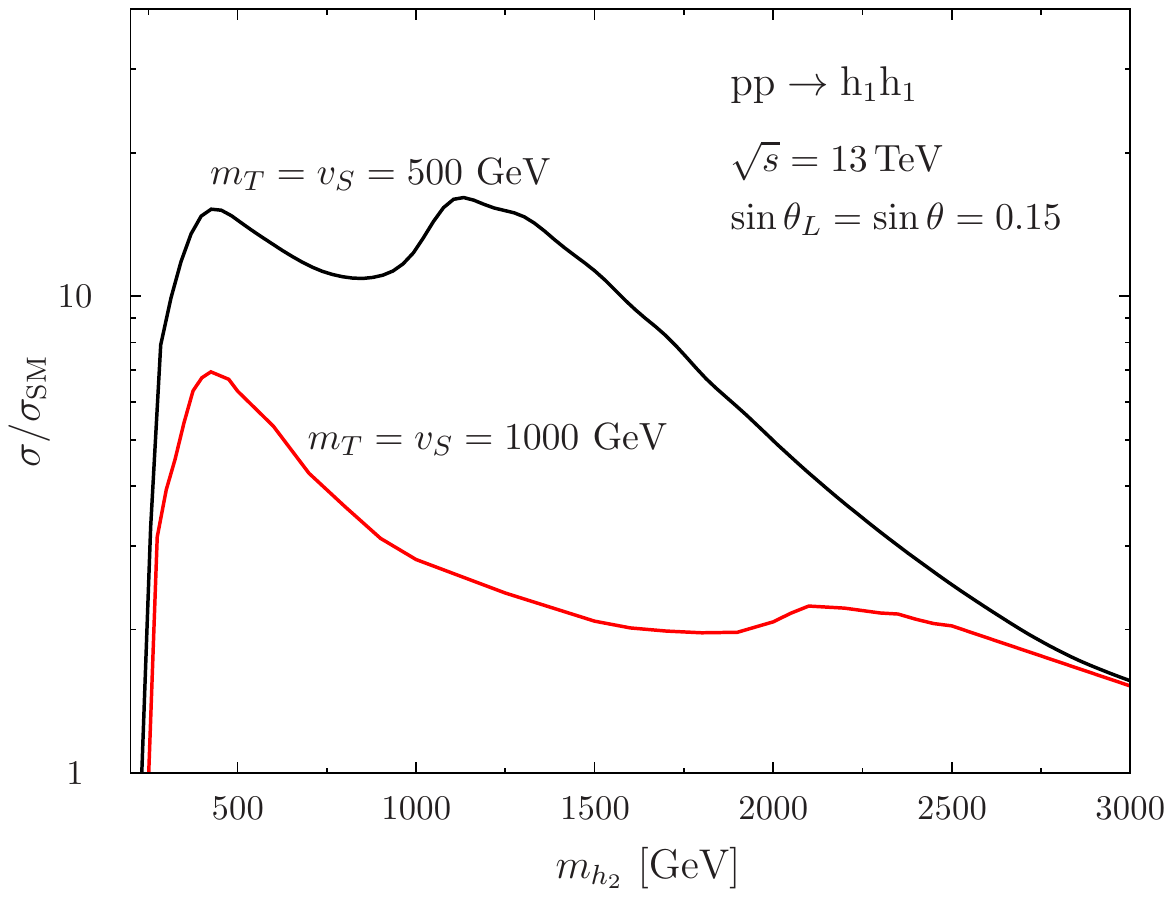}}
\vspace*{-0.0cm}
\caption{(Left) The 125 GeV Higgs plus jet cross section, $pp \to h_1 + {\rm jet}$, as a function of a cut on the $p_\perp$ of the Higgs for $\sqrt s =  13$~TeV. Shown is the ratio of the simplified Higgs model and the SM prediction, both at leading order, for different values of $m_T$. (Right) Higgs pair production, $pp \to h_1 h_1$, at 13~TeV, relative to the SM prediction, for two different values of $m_T$ and $v_S$. In both panels we have set $\sin\theta_L =  \sin\theta = 0.15$.}
\label{fig:h1_prod}
\end{figure}

Let us first consider the Higgs plus jet cross section, $pp \to h_1 + {\rm jet}$, as a function of a cut on the $p_\perp$ of the Higgs, Fig.~\ref{fig:h1_prod} (left). The calculation has been performed with {\tt MadGraph5${}_{}$aMC@NLO}~\cite{Alwall:2014hca,Hirschi:2015iia}, with the default settings for the parton distribution functions and scales. We 
have set $\sin\theta_L =  \sin\theta = 0.15$ (which is at the outer envelope of the allowed parameter space from the fit, and thus provides an upper bound on how large deviations from the SM can be), and consider two choices for the mass of the vector-like $T$ quark, $m_T = 500$\;GeV and $m_T = 1$\;TeV, respectively. The lower of these masses for $m_T$ is naively ruled out by current LHC searches, however in models such as ours with allow for dilution of the standard $T$ branching ratio the limits may be lower. In the case our choice of $m_T=500$~GeV, similar to the choices for $\sin\theta$ and $\sin\theta_L$ allows us to estimate the maximum size of the deviations away from SM behaviour.

While the emission of a hard jet allows for the exploration of the structure of the heavy fermion loop in principle, see {\it e.g.},~\cite{Azatov:2013xha,Grojean:2013nya,Schlaffer:2014osa,Buschmann:2014twa,Dawson:2015gka}, we see that the deviations from the SM prediction, which ranges from $1.63$\;pb for $p_\perp > 100$\;GeV to $4.5$\;fb  at  $p_\perp > 500$\;GeV, turn out to be numerically small in our model, given the allowed range of masses and mixings.  It will thus be challenging to constrain the properties of the vector-like $T$ quark from a measurement of the Higgs+jet cross section. Another way to probe new physics in the Higgs coupling to gluons is through double Higgs production~\cite{Dolan:2012ac,Contino:2012xk,Dawson:2015oha,Azatov:2015oxa}. We show the corresponding cross section, $pp \to h_1 h_1$ in Fig.~\ref{fig:h1_prod} (right). Here, we have again set $\sin\theta_L =  \sin\theta = 0.15$, and explore the cross section as a function of the mass of the vector-like $T$ quark, $m_T$, the singlet vacuum expectation value, $v_S$, and the mass of the heavy scalar, $m_{h_2}$. We find potentially sizable deviations from the SM cross section, $\sigma_{\rm SM} = 14.6$\;fb, resulting from the resonant production $pp \to h_2 \to h_1 h_1$. The size of the cross section depends not only on the mass of the heavy scalar, but also on the vacuum expectation value $v_S$ and the mass of the vector-like quark, which leads to a second threshold at $m_{h_2} \sim 2 m_T$. The observation of double Higgs production would thus be important to probe new physics in the loop-level coupling of the Higgs to gluons. We further note that previous work has focussed on the cases where there is either a top partner, or a new dihiggs resonance, but in general not both simultaneously. 
	
\begin{figure}[tb]
	\centerline{\includegraphics[width=0.75\textwidth,angle=0]{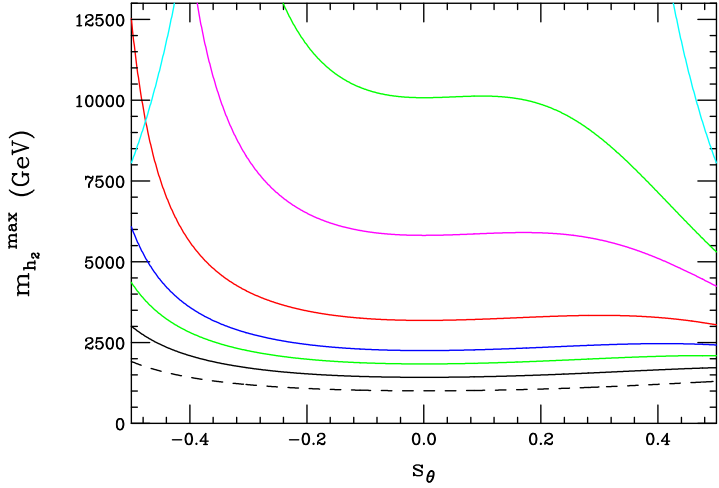}}
	\vspace*{-1.0cm}
	\caption{Constraints on the maximum value of the $h_2$ scalar mass from perturbative unitarity of the $h_2$ 
	self-scattering amplitude as a function of the $h-S$ mixing $\sin\theta$.
	From top to bottom (on the right-hand side of the plot) the curves correspond to values of $r\equiv v_H/v_S$ of 0.0, 0.01, 0.03, 0.1, 0.2, 0.3, 0.5 and 1.0.  The allowed region of parameter space is below the relevant constraint line.}
	\label{fig:Spert}
\end{figure}

We now examine the production and decay of the heavy scalar $h_2$ in more detail. 
Perturbative unitarity considerations can be used to constrain the mass of $h_2$ as a function 
of the input parameters $r\equiv v_H/v_S$ and the $h-S$ mixing angle $s_\theta$, once the mass of $h_1 \simeq h$, $\sim 125$ GeV is known. To see this, we begin by considering Eq.(47) in Ref.~{\cite{Xiao:2014kba}} for the $h_2$ self-scattering amplitude,
$h_2h_2\to h_2h_2$, which we can write as
\begin{equation}
{\cal M} = {1\over {64v_H^2}}~F(r,s_\theta,m_{h_2})\,,
\end{equation}
where we define $F(r,s_\theta,m_{h_2}) = a(r,s_\theta) m_{h_2}^2 + b(r,s_\theta) m_{h_1}^2$ with
\begin{eqnarray}
a(r,s_\theta) & = & 30(1+r)-45(1-r)c_2+18(1+r)c_4-3(1-r)c_6+24s_2^3\sqrt r\,, \\ 
b(r,s_\theta) & = & 6(1+r)-3(1-r)c_2-6(1+r)c_4+3(1-5)c_6-24s_2^3\sqrt r\,. \nonumber
\end{eqnarray}
Here we have employed the notation $c_n=\cos n\theta$. Since perturbative unitarity requires that 
$|{\cal M}| <1/2$, we 
can easily derive a bound on $m_{h_2}$ as a function of the input parameters.  By using the relation $\sqrt 2 G_F=1/v_H^2$ and defining $d=512\pi/\sqrt 2 G_F \simeq (9.8749 ~\rm TeV)^2$ we obtain:
\begin{equation}
m_{h_2} < m_{h_2}^{\rm max}={{(d-bm_{h_1}^2)}/{a}}\,.
\end{equation}
Numerical results using this expression are shown in Fig.~\ref{fig:Spert},  where we see several interesting effects: ($i$) For small values 
of $r$ the constraints are rather weak, particularly if $|s_\theta|$ also takes on small values. However as $r$ increases the 
value of $m_{h_2}^{\rm max}$ lies within a range that may be accessible via direct production at the LHC, although probing more of the parameter space would require a higher energy hadron collider. 
($ii$) Due to the last terms in the expressions for $a$ and 
$b$ which are proportional to $\sim s_2^3$, the bound is a not an even function of $s_\theta$ and displays a rather complex $h-S$ mixing angle 
dependence if $r$ is not too large.  Likewise the mixing angle dependence of the constraints flattens out 
when $r$ reaches values $\sim 0.2-0.3$ or greater. ($iii$) 
When $s_\theta \to 0$ the bound simplifies greatly, as in this limit $a\to 96r$ and $b\to 0$ such that we obtain 
$m_{h_2}^{\rm max}\simeq 1.01~{\rm TeV}/\sqrt r$. 

\begin{figure}[t!]
		\centerline{\includegraphics[width=0.62\textwidth,angle=0]{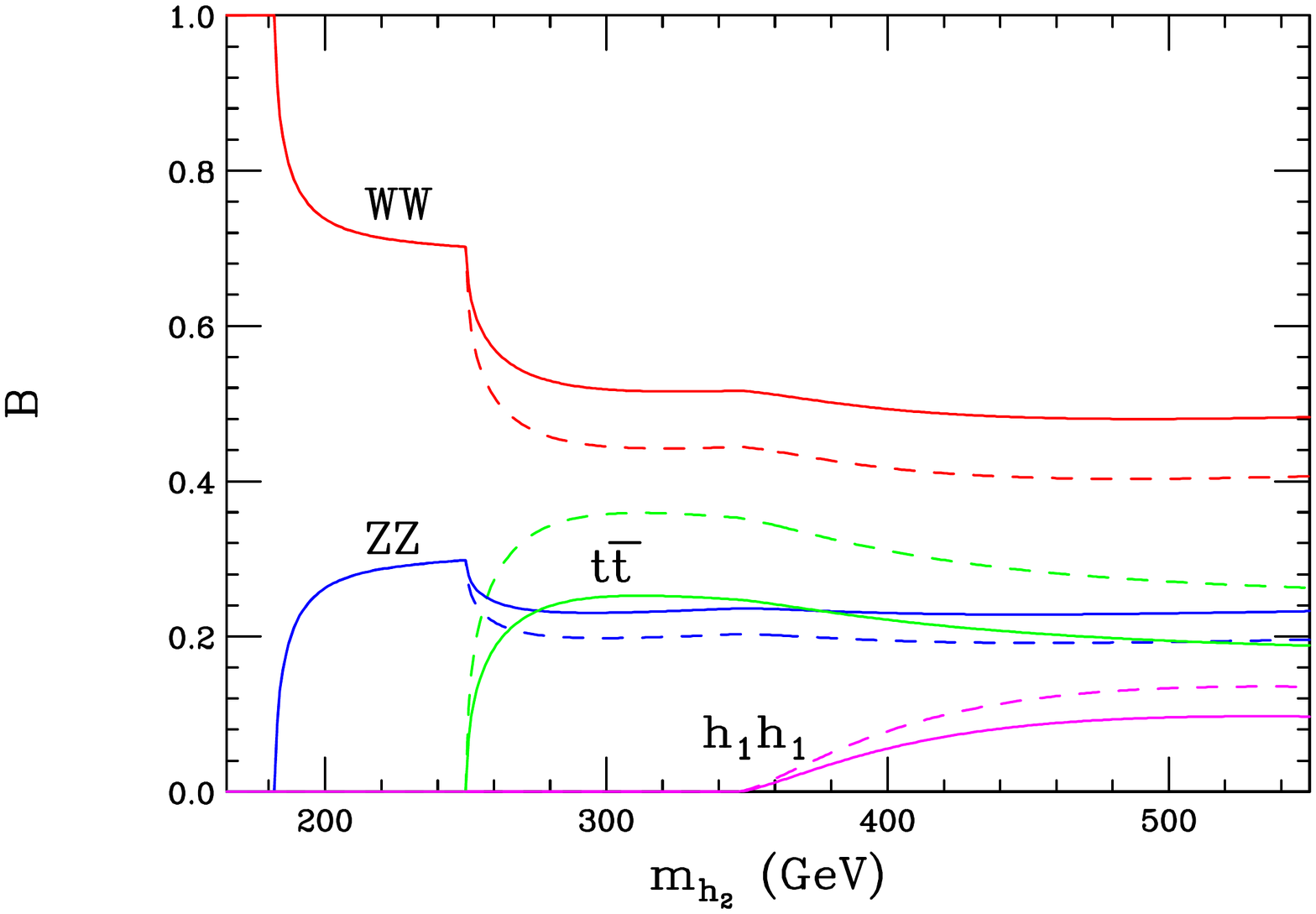}
			\hspace*{-2.0cm}
			\includegraphics[width=0.62\textwidth,angle=0]{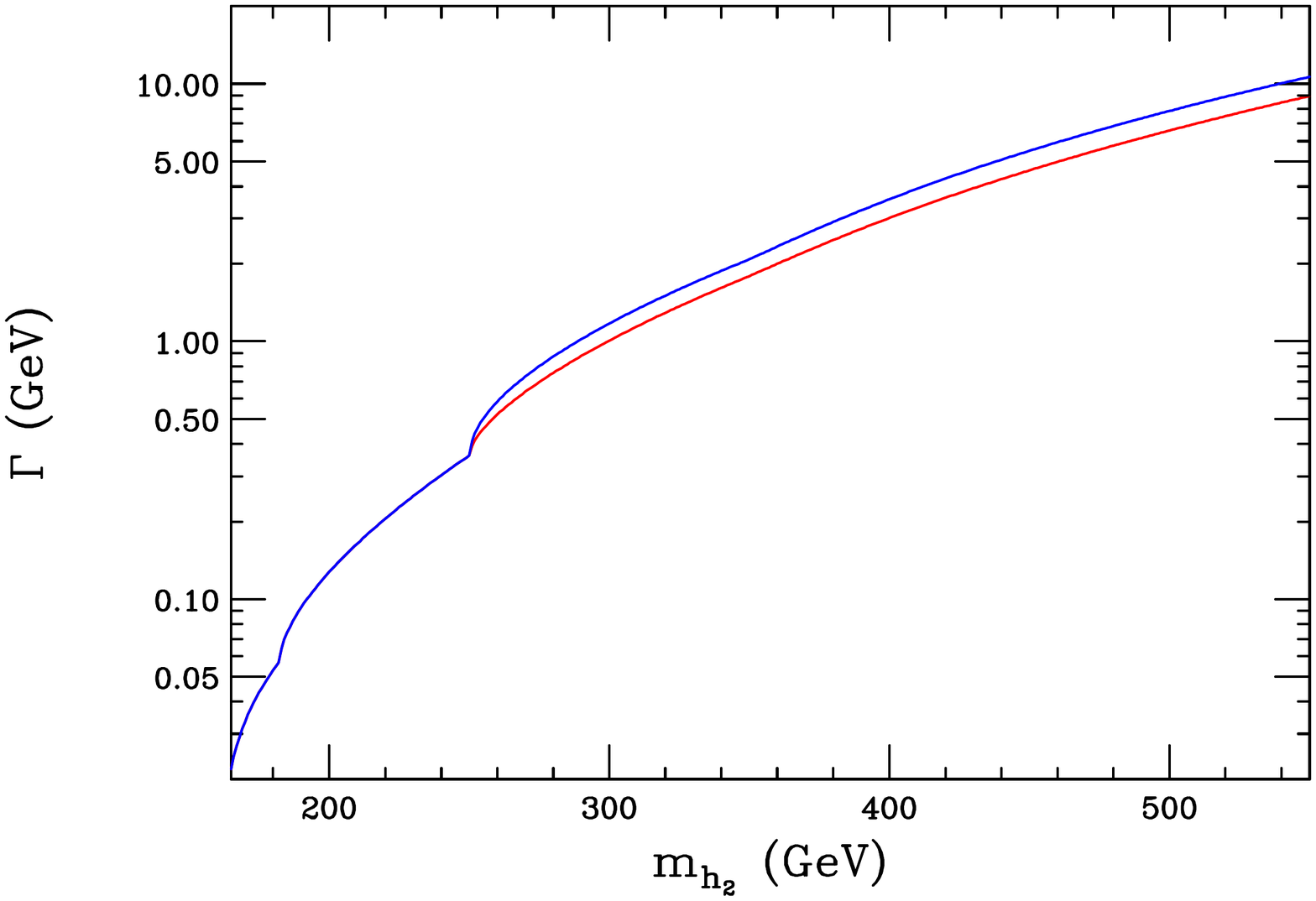}}
		\vspace*{-1.0cm}
		\caption{(Left) Branching fractions for $h_2$ as a function of its mass for $W^+W^-$(red), $ZZ$(blue), $t\bar t$(green) 
			and $h_1h_1$(magenta) final states, assuming that decays into the $T$-quark are kinematically inaccessible. Decays to $b\bar b$ and lighter states 
			have been neglected. Here we assume $v_S=500$ GeV, $s_L=0.15$ and $s_\theta=-0.15(0.15)$, corresponding to the dashed(solid) curves. (Right) The total $h_2$ decay width as a function of its mass in these two scenarios with the blue(red) curve corresponding to the case of $s_\theta>0(s_\theta<0)$.}
		\label{S-Decay}
	\end{figure}

	For completeness one might wonder if the sign of $s_\theta$ also has an important influence on the decays of $h_2$ itself. 
	Fig.~\ref{S-Decay} shows the corresponding $h_2$ decay branching fractions for two (different) model points which are 
	identical except for flipping the sign of $s_\theta$. The most obvious difference due to the sign in this example is the partial suppression of the gauge modes in favor of the light Higgs and top-quark final states when the sign of $s_\theta$ is flipped, although the effects are relatively mild. We also see from the right-hand panel of this figure that the $h_2$ total width is not much influenced by this choice of sign for these fixed values of the other model parameters.  

\begin{figure}[tb]
	\centerline{\includegraphics[width=0.62\textwidth,angle=0]{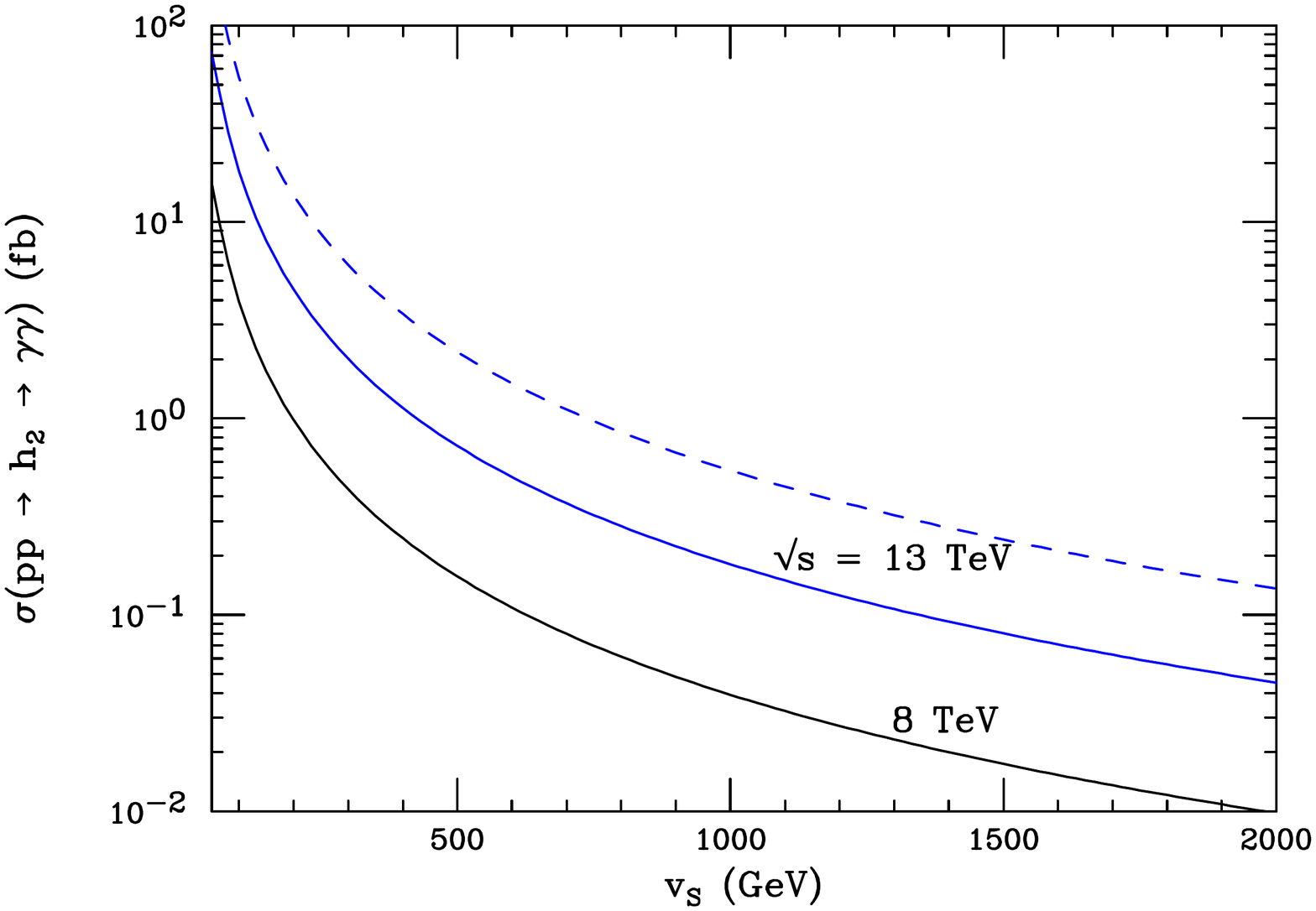}
		\hspace*{-2.0cm}
		\includegraphics[width=0.62\textwidth,angle=0]{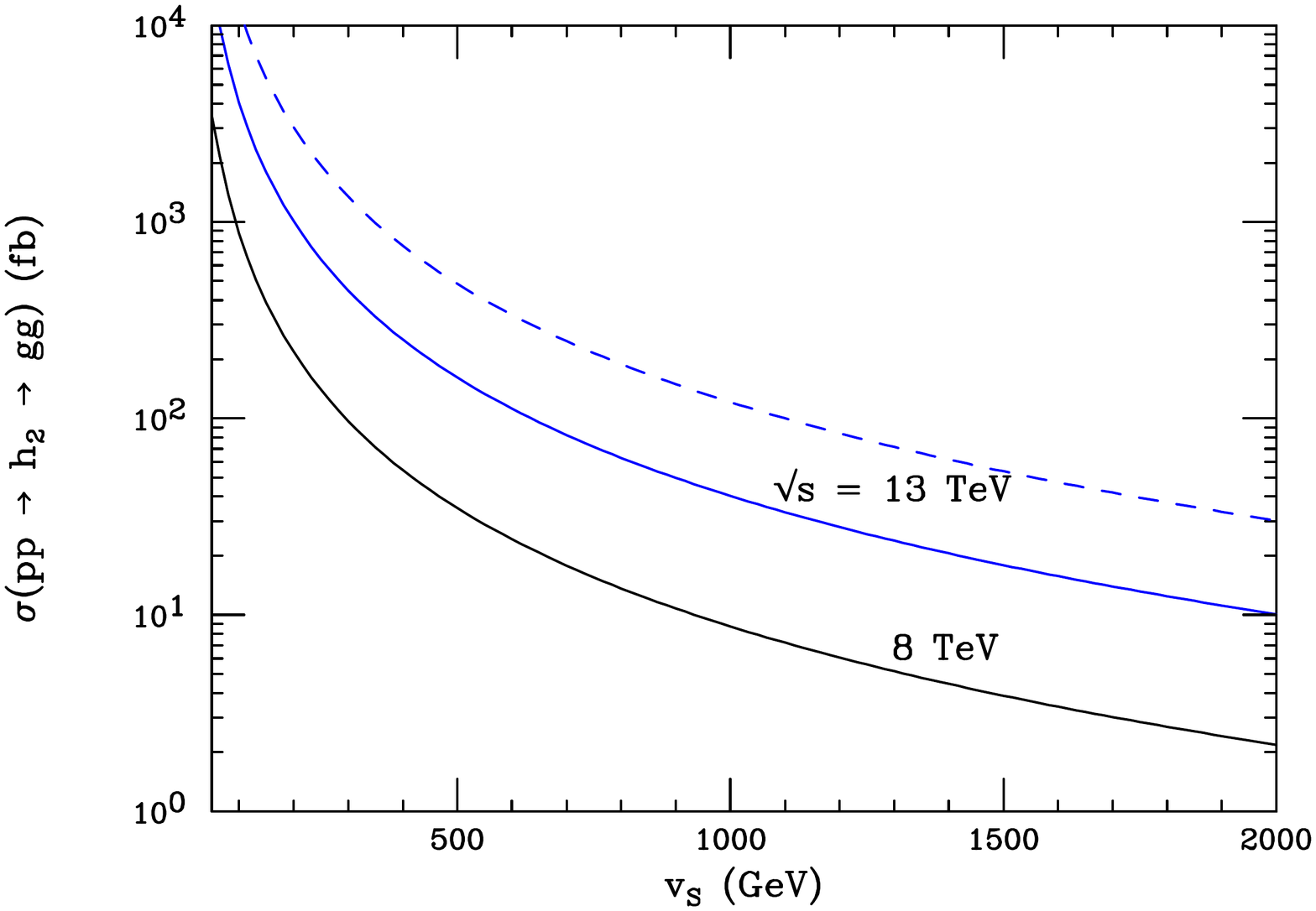}}
	\vspace*{-1.0cm}
	\caption{(Left) The cross-sections at 8 and 13~TeV for resonant $pp\to h_2 \to \gamma\gamma$ production as a function of the singlet vev $v_s$.  (Right) The cross-sections at 8 and 13~TeV for resonant $pp\to h_2 \to jj$ production as a function of the singlet vev $v_s$.   Here, mixing has been neglected as discussed in the text and $m_T=1 TeV$ is assumed. The solide blue (black) curve represents $\sqrt s = 13, 8$ TeV, while the dashed blue curve corresponds to the case with 3 generations of $T$-quarks contributing to the production at $\sqrt s = 13$ TeV. }
	\label{fig:diphotons_jets}
\end{figure}

Finally, we discuss the possibility of this simplified model as an explanation for the possible excess of diphoton events at 750~GeV recently reported by the ATLAS and CMS Collaborations in their initial $\sqrt{s}=13$~TeV data set~\cite{ATLAS-CONF-2015-081,CMS-PAS-EXO-15-004}. In order to reproduce this observation, a diphoton resonance with a cross-section of a
few inverse femtobarns
is required~\cite{Falkowski:2015swt,Buckley:2016mbr}. This cross-section assumes a narrow resonance, and includes the constraints from the 8~TeV diphoton searches. While the ATLAS data suggests a relatively large width of $\Gamma=40-50$~GeV, the CMS results are consistent with the new state being narrow, yet consistent with a wide resonance at the $2\sigma$ level.  The work of~\cite{Buckley:2016mbr} shows that a global fit to the 8 and 13 TeV data slightly prefers a narrow resonance.

A number of theoretical studies have suggested that this could be explained by the presence of a singlet scalar resonance and heavy vector-like fermions~\cite{Ellis:2015oso,Falkowski:2015swt,McDermott:2015sck,Benbrik:2015fyz,Gupta:2015zzs,Han:2015dlp,Knapen:2015dap,Zhang:2015uuo,Craig:2015lraxo,Altmannshofer:2015xfo}.  Such a resonance is required to be mostly-singlet in order to suppress tree-level decays to massive SM fermions and vector bosons. We show in the left panel of Fig.~\ref{fig:diphotons_jets} the cross-sections for $pp\to h_2 \to \gamma\gamma$ at 8 and 13~TeV for $m_T=1$~TeV assuming no mixing, {\it i.e.}, $s_\theta\sim 0$ and $s_L\sim 0$.  The solid blue (black) curves correspond to $\sqrt s = 13 (8)$ TeV.  The K-factor is set to be 2.0 according to~\cite{Harlander:2012pb,Harlander:2005rq,Harlander:2002wh}.  We set the $h-S$ mixing to zero for simplicity, as it will not have much influence on the production rate and this easily prevents $h_2$ decays into the electroweak gauge bosons $W$ and $Z$.  The quark mixing is set to vanish in order to avoid $t\bar t$ decays of the $h_2$ which are not observed, and thus increases its diphoton branching fraction. Fitting the observed cross-section requires $v_s$ to take on values of the order of 100 GeV.  This borders on exceeding the perturbative unitarity constraints discussed earlier, a fact which has already been noted in the literature by some. The blue dashed curve in this figures represents the case where there are 3 degenerate generations of $T$-quarks, which all contribution at loop-level to $h_2$ production and decay.  We see that in this case, perturbative values of the vev $v_S$ allow for a production rate that is consistent with observations.

The production of the $h_2$ resonance in gluon fusion also implies that there must be a corresponding dijet signal. In the right-hand panel of Fig.~\ref{fig:diphotons_jets} we show the $pp\to h_2 \to jj$ cross-section at 8 and 13~TeV. It is clear that current LHC dijet analyses do not yet constrain this parameter space, although this may change in the near future through trigger-level analyses. We note that the $h_2$ resonance is narrow with a width always less than a GeV.

%%%%%%%%%%%%%%%%%%%%%%%%%%%%%%%%%%%%%%%%%%%%%%%%%%%%%%%%%%%%
	\section{$T$ Production and Decay at the LHC}
	\label{sec:tprime}
%%%%%%%%%%%%%%%%%%%%%%%%%%%%%%%%%%%%%%%%%%%%%%%%%%%%%%%%%%%%%5	

In this section we explore the phenomenology of the heavy vector-like $T$-quark.  Before turning our attention to its
production rate and decay modes, we first examine the constraints derived from perturbative unitarity considerations of the 
$T$ self-scattering amplitude.
	
\begin{figure}[tb]
	\centerline{\includegraphics[width=0.65\textwidth,angle=90]{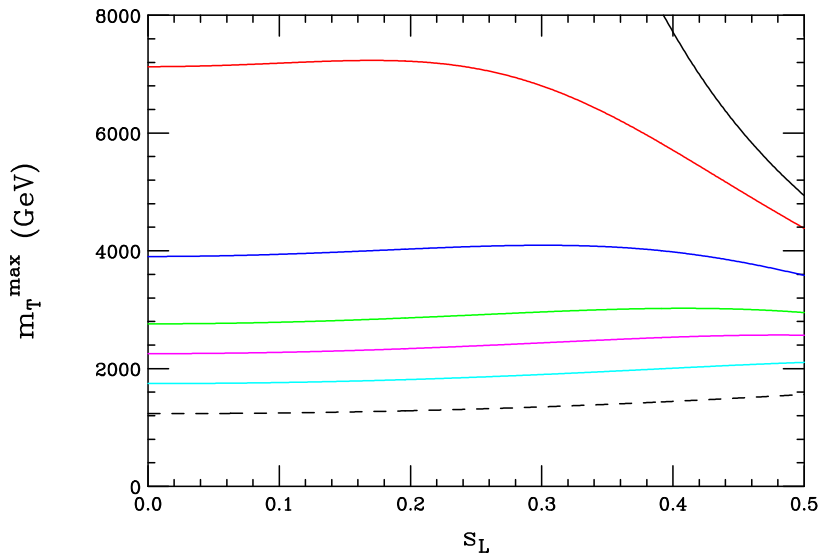}}
	\vspace*{-1.0cm}
	\caption{Constraints on the maximum value of the $T$-quark mass from perturbative unitarity of the $T$ self-scattering amplitude.
	From top to bottom (on the right-hand side of the plot) the curves correspond to values of $r\equiv v_H/v_S$ of 0.0, 0.03, 0.1, 0.2, 0.3, 0.5 and 1.0. The allowed region of parameter space is below the relevant constraint line.}
	\label{fig:Tpert}
\end{figure}

In a manner similar to the case of $h_2$ discussed above, the mass of the $T$-quark can also be constrained as a function of $r\equiv v_H/v_S$ and 
$s_L$ by employing perturbative unitarity requirements. To see this, we employ 
Eqs.(49)-(51) for the $T\overline{T}$ scattering amplitude in 
Ref.~{\cite{Xiao:2014kba}} and require $|{\cal M}|<8\pi$. From this we obtain the constraint
\begin{equation}
m_T < m_{T}^{\rm max}= \Bigg [ {{8\pi}\over {\sqrt 2 G_F}}\Bigg ]^{1/2} \Big [r(1-s_L^2)^2+s_L^4 \Big]^{-1/2}\,.
\end{equation}
This bound is explored numerically in Fig.~\ref{fig:Tpert}, where we note several points: ($i$) $m_{T}^{\rm max}$ is an even function of $s_L$ 
and is relatively independent of $s_L$ in the range of interest once $r <0.3$. ($ii$) If $r>0.2$, $m_{T}^{\rm max}$ lies within 
the range accessible to the LHC. ($iii$) Furthermore, once $s_L$ is relatively small the bound simplifies to 
$m_{T}^{\rm max}\simeq 1.23~{\rm TeV}/\sqrt r$. These unitarity bounds demonstrate that a 100~TeV collider is likely required to fully probe the properties and existence of the top partner. 

	\begin{figure}[t!]
		\centerline{\includegraphics[bb = 100 450 520 710, clip, width=0.475\textwidth,angle=0]{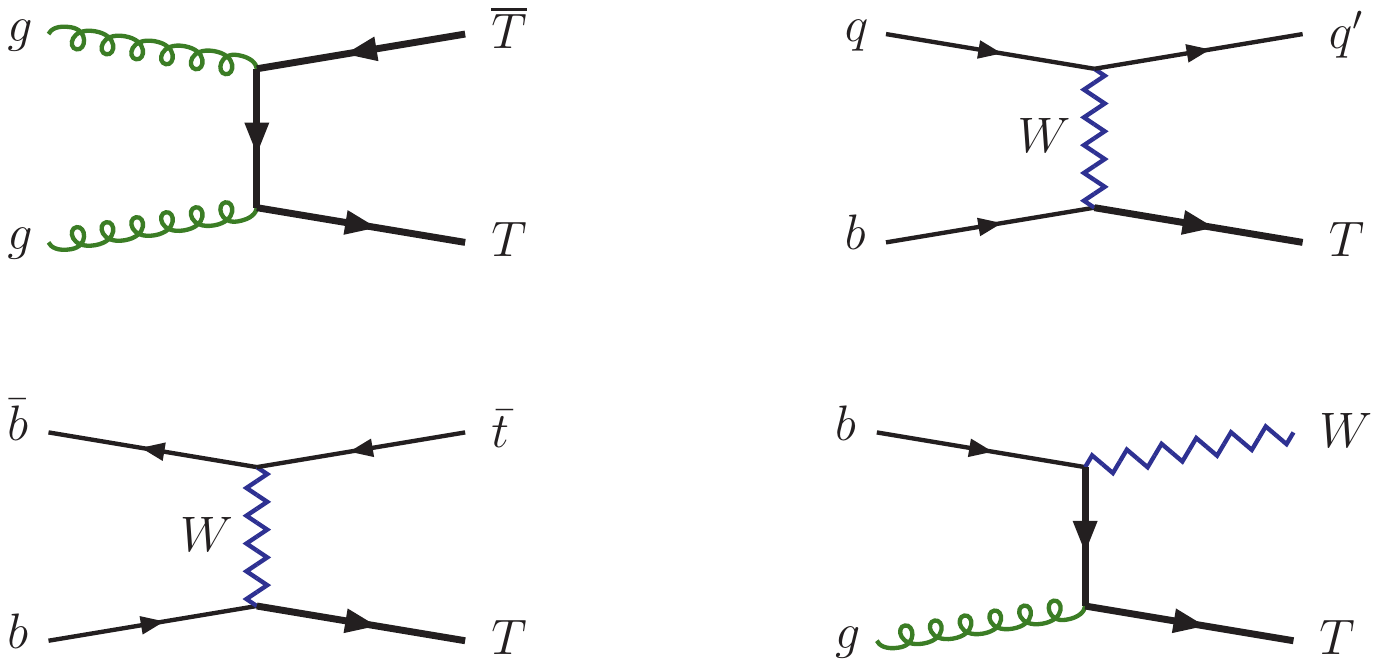}
			\hspace*{0.5cm}
			\includegraphics[bb = 150 535 500 800, clip, width=0.475\textwidth,angle=0,angle=0]{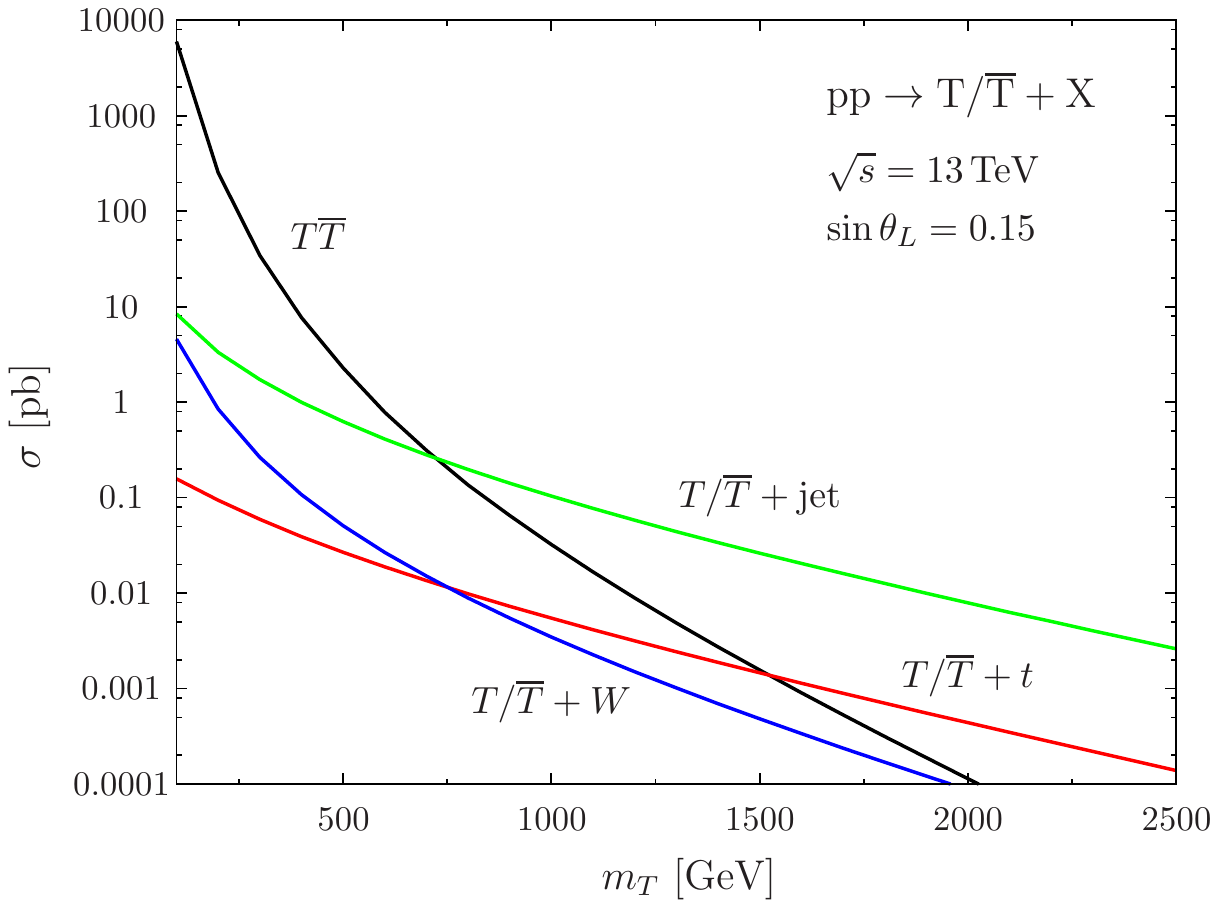}}
		\vspace*{00cm}
		\caption{(Left) Exemplary Feynman diagrams for $T\overline{T}$ (top-left) and single $T$ production at the LHC. (Right) The $gg, q\bar q \to T\overline{T}$ production cross section at 13 TeV (black) and the corresponding single $T$ production by several mechanisms: 
			$qb\to q'T$ (\ie, $T+$jet in green), $gb\to WT$  (\ie, $T+W$ in blue) and $q\bar q, b\bar b\to Tt$ (\ie,  $T+t$ in red). 
			We have taken $s_L=0.15$.  }
		\label{fig:fctp}
	\end{figure}

	\subsection {Production}

	As is the case for any new color-triplet fermion, $T\overline{T}$ production proceeds at leading order in QCD via both $q\bar q$ and 
	$gg$ annihilation, analogous to $t\bar t$ production in the SM. We show an example Feynman diagramn for the $gg \to T\overline{T}$ process in Fig.~\ref{fig:fctp} left in the upper-left hand corner. This production rate is quite large even for $T$ masses of order $\sim 1$ TeV as can be seen in the right-hand panel of Fig.~\ref{fig:fctp} which shows the cross-section at 13~TeV as a function of the $T$-quark mass. Resonant $T\overline{T}$ production can also occur through gluon fusion $gg \to h_2 \to  T\overline{T}$ when $2 m_T \leq m_{h_2}$. Given the current limits on $m_T$, this is unlikely to be relevant at the LHC, although it may be observable at a future higher-energy hadron collider.

        Due to the potentially significant $t-T$ mixing within the present scenario, somewhat more interesting single-$T$ production processes can also occur; examples of these production mechanisms are depicted in the left-hand panel of Fig.~\ref{fig:fctp}. 
	It is clear that single $T$-quark production can occur via several channels, in analogy to single top production in the SM. This can occur in a variety of ways, for example via $t$-channel $W$-exchange, $qb\to q'T$, which corresponds to single top production in the SM except that now the $T$ is more massive than the $t$ and the coupling, as discussed above, is suppressed by an additional factor of $s_L$. Similarly, $W+T$ associated production, $gb\to WT$, can occur via 
	$s$-channel $b$-exchange as well as $t$-channel $T$-exchange which is again analogous to single $t$ production in the SM 
	except for the larger $T$ mass and the amplitude-level suppression by a factor of $s_L$. Taking $s_L=0.15$ as a large, but allowed, value of the mixing given our above fit, Fig.~\ref{fig:fctp} shows that these cross sections can be quite significant, particularly for large $T$ masses~\cite{Aguilar-Saavedra:2013qpa,Ortiz:2014iza}. 
	We find that  for $m_T \gtrsim 750$ GeV the $T+$jet single production rate is larger than that for pair production from QCD due to 
	the significant phase-space suppression.  For other values of $s_L$, the cross sections for these processes can be easily obtained from the figure by rescaling by a factor of $(s_L/0.15)^2$. Finally, the single $T$ production process 
	$q\bar q, b\bar b \to \bar tT+h.c.$ can also now occur via $t$-channel $W$-exchange as well as $s$-channel flavor-changing  
	$Z$-exchange which leads to the smallest rate shown in the figure. Recall that this flavor-changing coupling arises from the fact that $t_L$ and $T_L$ have different values of the third component of weak isospin, $T_{3L}$. In this case the cross section scales as 
	$(s_Lc_L)^2$.
	
	Of course $t\overline{T}+h.c.$ production can also occur from both $gg$ and $q\bar q$ initial states via QCD through a loop-induced flavor-changing $(t\overline{T}+T\bar t)g$ coupling. Such an interaction vertex will be not only loop-suppressed but will also  involve the familiar factors of the $t-T$ mixing angle, $s_L$, leading to an even further suppression, and we do not consider it further in this work.

%%%%%%%%%%%%%%%%%%%%%%%%%%%%%%%%%%%%%%	
	\subsection{Decays}
%%%%%%%%%%%%%%%%%%%%%%%%%%%%%%%%%%%%%%	
		
	$T-t$ mixing allows for the decay of the $T$-quark and without which $T$ would be stable.  Within the present framework, any discussion of $T$ decays (and the collider searches for $T$) necessarily involves 
	the role of the additional mostly-singlet Higgs field, $h_2$. The tree-level decay modes of $T$ (if they are 
	kinematically allowed) are $T\to Wb, Zt, h_1t$ and $h_2t$, all of whose corresponding partial widths are proportional to the factor $s_L^2$.{\
          The presence of this `exotic' $h_2t$ decay mode, which is absent in almost all discussions of vector-like 
	heavy quark decays, can have a strong potential impact on the searches for $T$-quarks that have been performed 
	so far at the LHC. At present, existing searches only consider the case where the SM is solely augmented by a heavy $T$-quark with no additional 
	$h_2$ field being present. Thus these analyses all make the common assumption that the sum of the branching fractions for the 
	$Wb,Zt$ and $h_1t$ decay modes must sum to unity and lower bounds on the $T$ mass have been obtained for different branching fraction weights~\cite{Aad:2015kqa,Khachatryan:2015oba}. Under this assumption, ATLAS and CMS obtain the 95$\%$ CL lower bound 
	of $m_T \gtrsim 715-730$ GeV with even larger values being obtained as the various $T$-quark branching fractions are scanned over. Since these standard searches rely on reconstructing the $T$ mass from the SM decay products, they have minimal sensitivity to the $T\to h_2 t$ decay channel (the exception to this would be if $m_{h_1} \sim m_{h_2}$ such that the acceptances would be similar). However, as we will see below, there can be a significant region of the model parameter space where the $h_2 t$ decay mode is kinematically allowed and obtains a respectable coupling strength relative to that for $h_1 t$. Clearly the influence of this mode on the lower bound obtained on 
	the $T$ mass will depend upon how $h_2$ itself decays. We might expect that if the $h_2$ decays in a manner broadly similar 
	to the SM-like $h_1$ ({\it e.g.}\, into $WW$, $ZZ$ or $b\bar b$) the effects will be minimal except that the final state kinematics 
	can be significantly different depending upon both the $T$ and $h_2$ masses. Of course if $h_2$ is sufficiently heavy, the 
	 branching ratio for the decay $h_2 \to t \bar t$ can be significant so that the decay path $T\to tt \bar t$ via virtual $h_2$ exchange opens up, this has also not been 
	examined in $T$-quark searches.  Clearly a detailed analysis of how the existence of a non-negligible branching fraction for 
	$T$ into the $h_2 t$ mode would affect the searches for $T$-quarks at the LHC remains an open question that needs to be performed in detail, but we might expect that if this branching fraction is sufficiently small the `standard' limits discussed above will approximately apply.  

	\begin{figure}[t!]
		\centerline{\includegraphics[width=0.59\textwidth,angle=0]{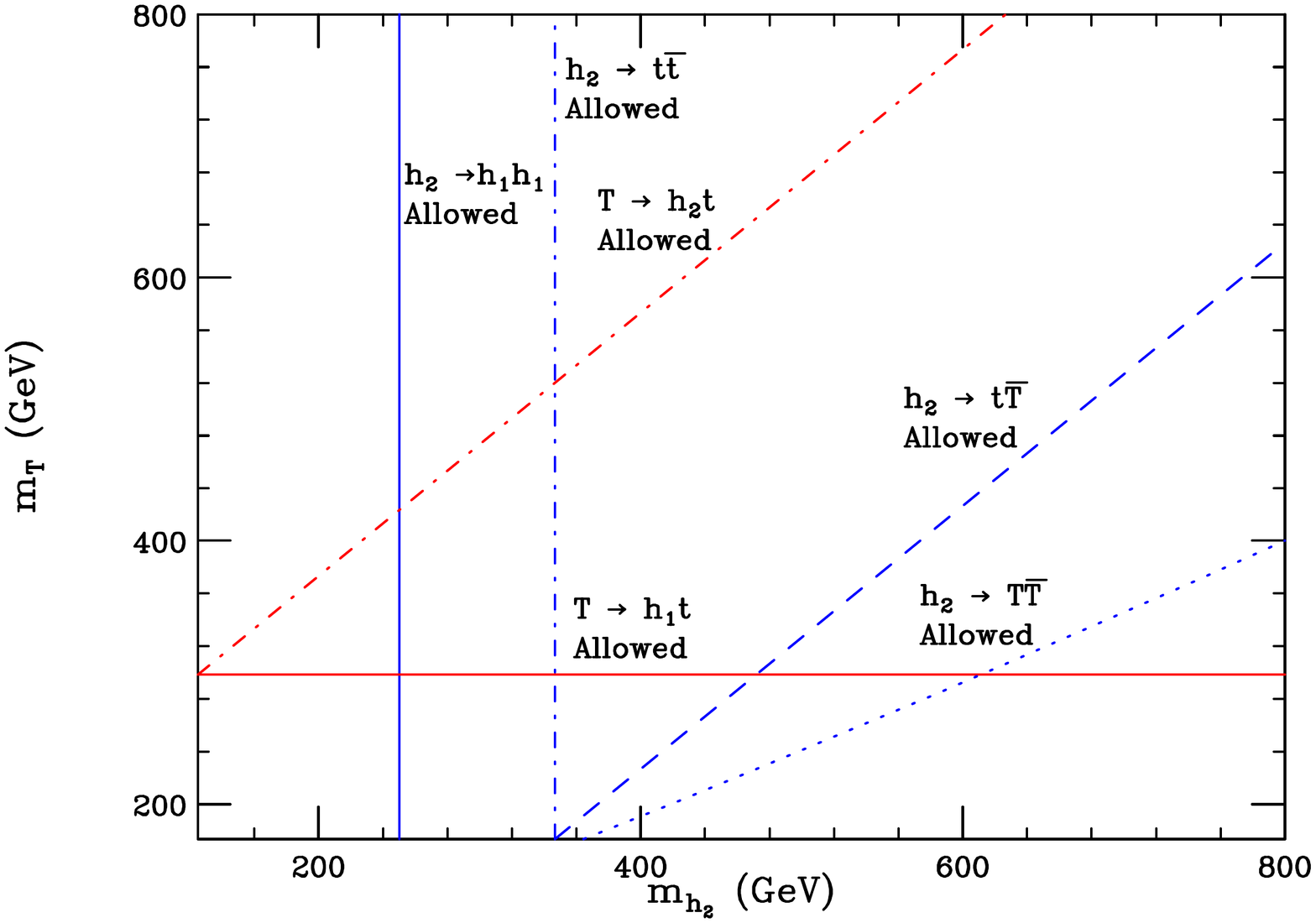}
			\hspace*{-2.0cm}
			\includegraphics[width=0.60\textwidth,angle=0]{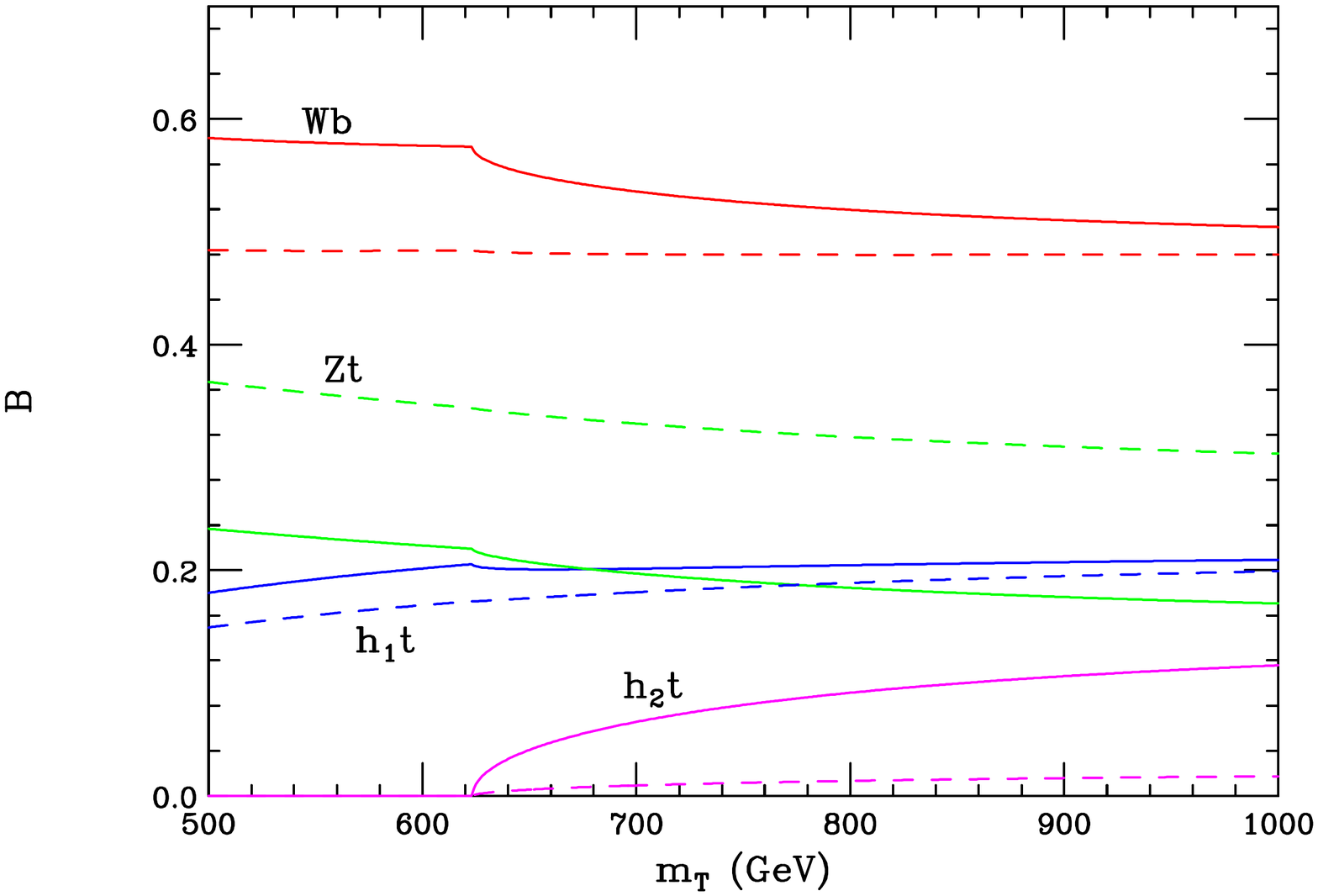}}
		\vspace*{-1.0cm}
		\caption{(Left) Kinematic regions for the various decays of $h_2$ and $T$ in the $m_{h_2}-m_T$ plane. To the right of the blue 
			solid (dash-dotted, dashed, dotted) lines the decay $h_2\to h_1 h_1(t\bar t,~t\overline{T}+h.c., T\overline{T})$ are kinematically allowed. 
			In the region above the red solid (dash-dotted) line the decay $T\to h_1 t(h_2 t)$ is kinematically allowed. (Right) Sample $T$ 
			decay branching fractions as a function of $m_T$ assuming $m_{h_2}=300$ GeV, $s_L=0.15$, $v_S=500$ GeV and 
			$s_\theta=-0.15(0.15)$ corresponding to the dashed (solid) curves. The branching fractions for the $Wb(Zt,~h_1t,~h_2t)$    
			modes are shown as the red (blue, green, magenta) curves.}
		\label{regions}
	\end{figure}

	To provide some overall understanding of the interplay between the masses of $h_2$ and $T$ and the corresponding decay 
	physics, the left-hand panel of Fig.~\ref{regions} gives a semi-quantitative feel for the regions in this space where the various 
	decay modes may occur. The {\it strengths} of the various couplings are determined by the three remaining model  parameters $s_\theta, s_L$ and the ratio of vev's $r=v_H/v_S$ and are given in the Appendix. An important example is 
	provided by the ratio of the $T\to h_2t$ and $T\to h_1t$ partial widths, $R$, which apart from phase space factors is given 
	by
	\begin{equation}
		R={{\Gamma(T\to h_2 t)}\over {\Gamma(T\to h_1 t)}} \sim \Bigg[ {{s_\theta-rc_\theta}\over {c_\theta+rs_\theta}}\Bigg]^2\,,
	\end{equation}
	where $s_\theta=\sin \theta$ and $c_{\theta}= \cos \theta$. Clearly, the parameter space region with somewhat larger $r$ values and 
	with $s_\theta<0$ can lead to a significant result for this ratio, thus  providing a good example of a situation where 
	the sign of $s_\theta$ is important. In Fig.~\ref{regions} above, we see that there is a region where $R>1$ 
	is obtained.

	The right-hand panel of Fig.~\ref{regions} compares the branching fractions for the various $T$-quark decay modes for two parameter choices which differ only in the choice of sign for $s_\theta$.  There can be some significant differences in these two cases:  apart from the obviously much larger value for the branching fraction of the $h_1t$ mode when $s_\theta>0$ we see that the branching fraction for $h_2t$ is larger in the $s_\theta<0$ case at the expense of that for $h_1t$ at larger values 
	of $m_T$. On the other hand, we also see that the branching fractions for the $Wb,Zt$ decay modes are relatively unaffected by the change of sign of $s_\theta$. In general, since $\kappa_g \simeq c_\theta-s_\theta r$, much of the parameter space where 
	the above ratio of decay widths is large is restricted by LHC data at some level since it simultaneously produces a too large value for $\kappa_g$.

        \begin{figure}
          \centering
        \includegraphics[width=0.5\textwidth,height=0.3\textheight]{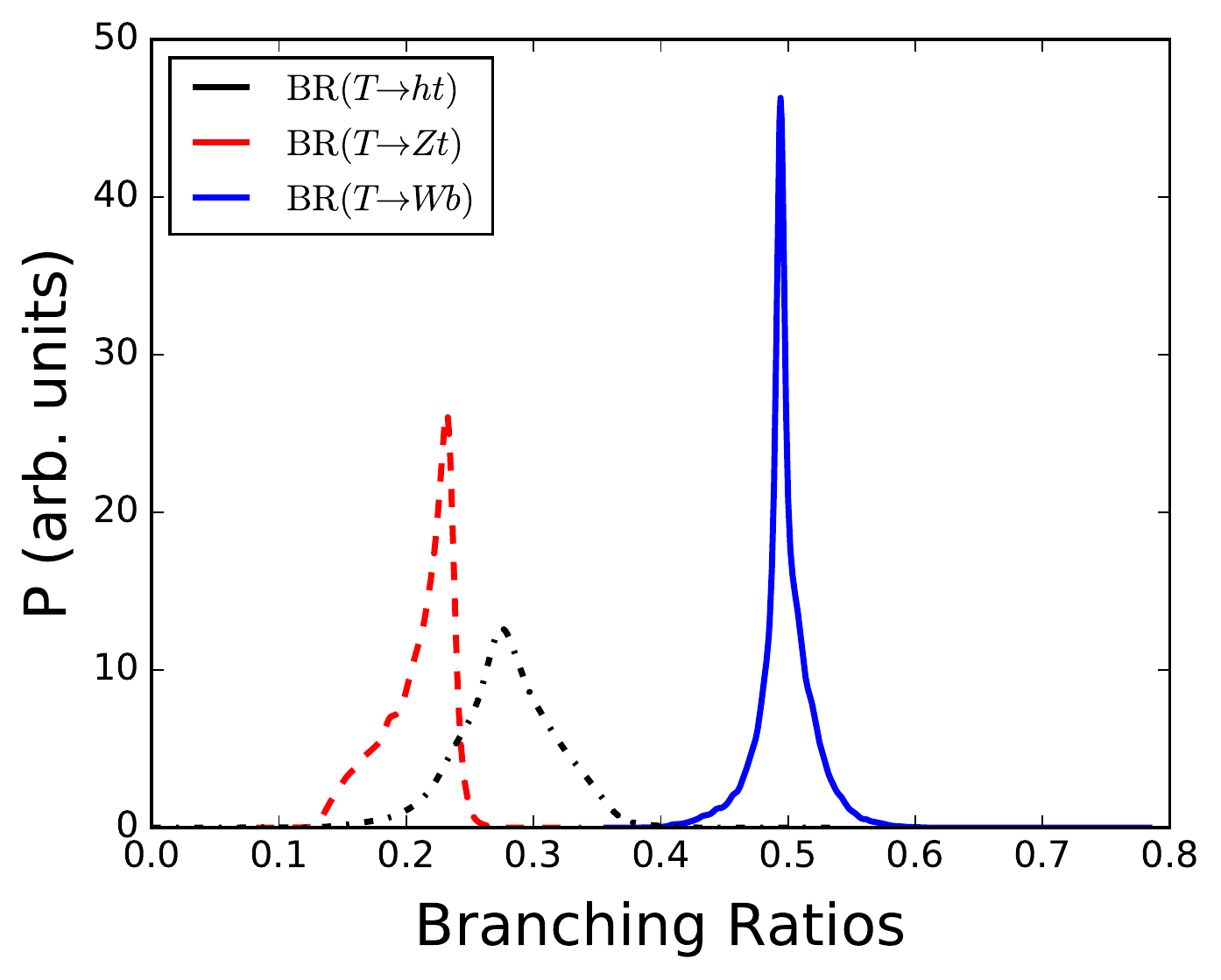}  
        \caption{This plot shows the probability distributions of the $\textrm{BR}(T\to h_1t)$ (black dash-dot), $\textrm{BR}(T\to Zt)$ (red dotted) and $\textrm{BR}(T\to Wb)$ (blue solid) branching ratios of the top-partner $T$ resulting from our fit to the Higgs and electroweak data. The $y$-axis shows probability in arbitrary units, and the areas of the three branching ratio curves have all been individually normalised to one. }
        \label{fig:BRT_plot}
        \end{figure}
        While this discussion describes  the phenomenology that is \textit{possible}, given the results of our fit, we can also ask about the \textit{probable} decays of $T$. We show in Fig.~\ref{fig:BRT_plot} the probability distributions for the branching ratios of $T$ decaying into SM final states: $T\to h_1t$, $T\to Zt$ and $T\to Wb$ in dot-dash black, dashed red and solid blue, respectively. We find that it is most likely given current data that the central values of the probability distribution for the branching ratios are close to the well-known ratios of $25:25:50$. However, there is substantial room for non-standard branching ratios with all decay channels subjection to variation on the order of 10\%. As discussed above, given the wide allowed ranges for the parameters, we also found it possible that the $\mathrm{BR}(T\to h_2t)$ maybe be as large as 10\%.

%%%%%%%%%%%%%%%%%%%%%%%%%%%%%%%%%%%%%%%%%
\section{Conclusion}
\label{sec:conc}
%%%%%%%%%%%%%%%%%%%%%%%%%%%%%%%%%%%%%%%%%%5

We have formulated a set of simplified models to characterize the interactions of physics beyond the Standard Model with the 125 GeV Higgs boson.  We selected one such model, where
 the Standard Model is extended by a gauge singlet scalar and vector-like fermions which mix with the SM top-quarks, and studied its phenomenology in detail.  In particular, we examined 
the complementarity between indirect searches for new physics in precise determinations of the 125 GeV Higgs couplings and distributions with direct searches for new particles.  We constrained the model parameters by performing a global fit using the ATLAS and CMS combined data on the Higgs couplings and searches for heavy new scalars. We argued that although our model allows for the possibility of exotic top partner phenomenology, it is robustly constrained by top partner searches from LHC Run 1.  Nonetheless, the top-partner decay into a SM top-quark and new heavy scalar, $T\to th_2$, would lead to novel signatures and deserves to be better explored with Run 2 searches in mind. We find that given current direct search constraints, the effects of the top partners in Higgs loops are mostly likely too small to observe, in line with general expectations in weakly coupled models. We also note that our model (at least in the case a single generation of top partners) is unable to explain a possible 750~GeV diphoton resonance which has recently been reported by ATLAS and CMS without resorting to non-perturbative couplings. As the LHC moves further into the precision era, simplified models for Higgs physics will serve as a test-bed for expectations of possible BSM signals in the Higgs sector and direct searches for new particles.

\section*{Acknowledgement}
We thank Tim Stefaniak for assistance with the ATLAS+CMS combination in HiggsSignals.
The research of MD was performed in part at the Munich Institute for Astro- and Particle Physics (MIAPP), part of the DFG cluster of excellence ``Origin and Structure of the Universe''. MK would like to thank SLAC and the SLAC Theory group for hospitality and support. The work of MK is supported by the German Research Foundation DFG through the research unit 2239 ``New physics at the LHC''.   The work of JLH and TGR was supported by the U.S. Department of Energy, Contract DE-AC02-76SF00515.

\appendix

\section{Model details}
\label{app:model}

\subsection{The scalar potential}\label{sec:scalar}

To minimize the potential in Eq.(\ref{eq:potential}), we need to solve 
\begin{equation}
\left. \frac{\partial V(h,s)}{\partial h}\right |_{\langle H \rangle = v_H/\sqrt{2}, \langle S \rangle = v_S} = \left. \frac{\partial V(h,s)}{\partial s}\right |_{\langle H \rangle = v_H/\sqrt{2}, \langle S \rangle = v_S}  = 0.
\end{equation}
For a minimum, we also need 
$\partial^2V/\partial s^2 > 0$, $\partial^2V/\partial h^2 > 0$, and $\partial^2V/\partial s^2 \, \partial^2V/\partial h^2 - (\partial^2 V/(\partial s \partial h))^2 > 0$. We find
\begin{eqnarray}
\frac{\partial V}{\partial h} & = &  -\mu^2(h + v_H) + \lambda(h + v_H)^3 \nonumber \\
&& + \frac{a_1}{2}(h + v_H)(s + v_S) +  \frac{a_2}{2}(h + v_H)(s + v_S)^2\,, \\
\frac{\partial^2 V}{\partial h^2} & = &  -\mu^2 + 3 \lambda (h+v_H)^2 + \frac{a_1}{2}(s + v_S) +  \frac{a_2}{2}(s + v_S)^2\,, \\[1mm]
\frac{\partial V}{\partial s} & = &  \frac{a_1}{4} (h+v_H)^2 + \frac{a_2}{2}(h+v_H)^2(s+v_S)\nonumber\\
&& + b_1 + b_2 (s+v_S) + b_3 (s+v_S)^2 + b_4 (s+v_S)^3\,, \\[1mm]
\frac{\partial^2 V}{\partial s^2} & = & \frac{a_2}{2}(h+v_H)^2 + b_2 + 2 b_3 (s +v_S) + 3 b_4 (s+v_S)^2\,, \\[1mm]
\frac{\partial^2V }{\partial s \partial h} & = & (h+v_H)\, \left (\frac{a_1}{2} + a_2(s+v_S)\right).
\end{eqnarray}
Thus the conditions for a minimum are (assuming $v_H, v_S > 0$): 
\begin{eqnarray}
-\mu^2 + \lambda v_H^2 + \frac{a_1}{2}  v_S +  \frac{a_2}{2} v_S^2& = & 0\label{eq:min1}\,, \\
 \frac{a_1}{4} v_H^2 + \frac{a_2}{2} v_H^2 v_S + b_1 + b_2 v_S + b_3 v_S^2 + b_4 v_S^3 & = & 0 \label{eq:min2}\,, \\
\lambda   & > & 0 \,, \\
\frac{a_2}{2}v_H^2 + b_2 + 2 b_3 v_S + 3 b_4 v_S^2 & > & 0\,, \\
2 \lambda v_H^2 \left(\frac{a_2}{2} v_H^2 + b_2 + 2 b_3 v_S + 3 b_4 v_S^2 \right) - v_H^2 \left(\frac{a_1}{2} + a_2v_S\right)^2 &  > &  0.
\end{eqnarray}

The physical masses of the two scalar particles are determined by the mass matrix 
\begin{equation}
V(h,s) \supset  \frac12 (h \; s)
\begin{pmatrix}
M_{11}^2 & M_{12}^2 \\
M_{12}^2 & M_{22}^2 
\end{pmatrix}
\begin{pmatrix} h \\ s \end{pmatrix}, 
\end{equation}
where 
\begin{eqnarray}
M_{11}^2 & = & \left. \frac{\partial^2 V}{\partial h^2}\right |_{\langle H \rangle = v_H/\sqrt{2}, \langle S \rangle = v_S}   \;\;\,=  \;2 \lambda v_H^2 +\frac{a_1}{2} v_S \,, \\
M_{22}^2 & = & \left. \frac{\partial^2 V}{\partial s^2}\right |_{\langle H \rangle = v_H/\sqrt{2}, \langle S \rangle = v_S}   \;\;\,=  \;\frac{a_2}{2} v_H^2 + b_2 + 2 b_3 v_S + 3 b_4 v_S^2\,, \\
M_{12}^2 & = & \left. \frac{\partial^2 V}{\partial s\partial h}\right |_{\langle H \rangle = v_H/\sqrt{2}, \langle S \rangle = v_S}  \,\, =\; \left(\frac{a_1}{2} +a_2 v_S\right) v_H.
\end{eqnarray}
The physical masses are 
\begin{equation}
m_{1,2}^2 = \frac12\left( M_{11}^2 + M_{22}^2 \mp \sqrt{(M_{11}^2 - M_{22}^2)^2 + 4 M_{12}^4} \right),
\end{equation}
where we assume that the lighter mass eigenstate corresponds to the SM-like Higgs boson with $m_1 = 125$\,GeV.  
The mass eigenstates $h_1, h_2$ are related to the fields $h,s$ through 
\begin{equation}
\begin{pmatrix}
h_1\\h_2
\end{pmatrix} = 
\begin{pmatrix}
\cos\theta & -\sin\theta \\
\sin\theta & \cos\theta 
\end{pmatrix}
\begin{pmatrix}
h \\ s
\end{pmatrix},
\end{equation}
with 
\begin{equation}\label{eq:theta}
\tan(2\theta) = \frac{2 M_{12}^2}{M_{22}^2 - M_{11}^2}.
\end{equation}
In the limit $v_S \gg v_H$, and setting $a_1 = b_1 = b_3 = 0$, the expressions for the masses and the mixing angle simplify:
\begin{eqnarray}
m_1^2 & = & 2 \lambda v_H^2 \left(1 - \frac{a_2^2}{4 \lambda b_4}\right) \,, \\
m_2^2 & = & 2 b_4 v_S^2 \left(1+ \frac{a_2^2}{4 b_4^2}\frac{v_H^2}{v_S^2} \right) \,, \\
\tan(2\theta) & = & \frac{a_2}{b_4} \frac{v_H}{v_S}.
\end{eqnarray}

\subsection{The Yukawa potential}\label{sec:yukawa}

Here we calculate the mass mixing between the top quark and the new top partner field $T$.
The mass terms have the form 
\begin{eqnarray}
\mathcal{L}_{\rm Yukawa} & \supset &  
(\bar{t}^{\rm int}_L \; \overline{T}^{\rm int}_L)
\mathcal{M}
\begin{pmatrix} t^{\rm int}_R \\ T^{\rm int}_R \end{pmatrix}\\
& = & 
(\bar{t}^{\rm int}_L \; \overline{T}^{\rm int}_L)
\begin{pmatrix}
y_t v_H / \sqrt{2} & \lambda_T v_H / \sqrt{2} \\
0 & y_T v_S 
\end{pmatrix}
\begin{pmatrix} t^{\rm int}_R \\ T^{\rm int}_R \end{pmatrix}. \label{eq:massmatrix}
\end{eqnarray}
The physical mass eigenstates $t_{L/R}$ and $T_{L/R}$, and the mixing angles $\theta_{L/R}$, are obtained from bi-unitary transformations, 
\begin{equation}
\begin{pmatrix}
t_{L/R}\\ T_{L/R}
\end{pmatrix} 
= \mathcal{U}_{L/R}
\begin{pmatrix}
t^{\rm int}_{L/R}\\ T^{\rm int}_{L/R}
\end{pmatrix},
\end{equation} 
with 
\begin{equation}
\mathcal{U}_{L/R}  = 
\begin{pmatrix}
\cos\theta_{L/R} & -\sin\theta_{L/R} \\
\sin\theta_{L/R} & \cos\theta_{L/R}
\end{pmatrix} .
\end{equation}
The mass matrix $\mathcal{M}$ is diagonalized according to
\begin{equation}\label{eq:mixing}
\mathcal{U}_L \mathcal{M} \mathcal{U}_R^\dagger = 
\begin{pmatrix}
m_t & 0 \\
0 & m_T 
\end{pmatrix} \equiv \mathcal{M}_{\rm diag}
\end{equation}
or equivalently
\begin{equation}
\mathcal{U}_L \mathcal{M} \mathcal{M}^\dagger \mathcal{U}_L^\dagger = \mathcal{U}_R \mathcal{M}^\dagger \mathcal{M} \mathcal{U}_R^\dagger = 
\mathcal{M}_{\rm diag}^2.
\end{equation}
We use 
\begin{eqnarray}
{\rm Tr}\,[\mathcal{M}\mathcal{M}^\dagger] & = & m_t^2 +  m_T^2 \quad {\rm and}\\
{\rm Det}\,[\mathcal{M}\mathcal{M}^\dagger] & = & m_t^2 m_T^2 , 
\end{eqnarray}
to derive the masses: 
\begin{equation}
m_{t/T}^2  =   \frac14\left( y_t^2 v_H^2 + \lambda_T^2 v_H^2 + 2 y_T^2 v_S^2  \mp \sqrt{(y_t^2 v_H^2 + \lambda_T^2 v_H^2 + 2 y_T^2 v_S^2)^2-8 y_t^2 v_H^2 y_T^2 v_S^2}\right), 
\end{equation}
where we identify the lighter mass eigenstate $m_t$ with the physical top-quark mass $m_t = 173.2$\,GeV. 
From the fact that the off-diagonal terms of $\mathcal{U}_L \mathcal{M} \mathcal{M}^\dagger \mathcal{U}_L^\dagger$ and $\mathcal{U}_R \mathcal{M}^\dagger \mathcal{M} \mathcal{U}_R^\dagger$ vanish, we obtain the mixing angles: 
\begin{equation}
\tan(2 \theta_L) =  \frac{-2\sqrt{2} \lambda_T v_H \, y_T v_S}{y_t^2 v_H^2 + \lambda_T^2 v_H^2 - 2 y_T^2 v_S^2 } \quad {\rm and} \quad
\tan(2 \theta_R) =  \frac{- 2 y_t \lambda_T v_H^2}{y_t^2 v_H^2 - \lambda_T^2 v_H^2 - 2 y_T^2 v_S^2}.
\end{equation}
Note that the two mixing angles $\theta_L$ and $\theta_R$ are not independent. Using Eq.(\ref{eq:mixing}) we find that 
\begin{equation}
\tan\theta_R = \frac{m_t}{M_T} \tan\theta_L \equiv \sqrt{r_t}  \tan\theta_L, 
\end{equation}  
or
\begin{equation}
\sin^2\theta_R = \frac{r_t\sin^2\theta_L}{1-(1-r_t)\sin^2\theta_L}\quad {\rm and} \quad \cos^2\theta_R = \frac{\cos^2\theta_L}{1-(1-r_t)\sin^2\theta_L}.
\end{equation}  
In the limit $v_S \gg v_H$ the expressions for masses and mixing angles simplify and read
\begin{eqnarray}
m_t^2  =  \frac12 v_H^2 y_t^2 \left(1 -  \frac{\lambda_T^2}{2 y_T^2} \frac{v_H^2}{v_S^2}  \right), &&  
m_T^2 = v_S^2 y_T^2 \left(1 +  \frac{\lambda_T^2}{2 y_T^2} \frac{v_H^2}{v_S^2}  \right),\\
\tan(2\theta_L) =  {\sqrt2}\frac{\lambda_T}{y_T}\frac{v_H}{v_S}, &&  
\tan(2\theta_R)  =  \frac{\lambda_T y_t}{y_T^2}\frac{v_H^2}{v_S^2},  
\end{eqnarray}
up to corrections of $\mathcal{O}(v_H^3/v_S^3)$.

Let us work out the couplings of the Higgs to the top and bottom quarks. We first look at the terms including top quarks only and express the Yukawa Lagrangian, Eq.(\ref{eq:yuk}), in terms of the physical states $t, T$: 
\begin{eqnarray}
\mathcal{L}_{\rm Yukawa} & \supset & y_T S\overline{T}^{\rm int}_L T^{\rm int}_R + y_t \overline{Q}^{\rm int}_L\widetilde{H} t^{\rm int}_R + \lambda_T \overline{Q}^{\rm int}_L \widetilde{H} T^{\rm int}_R\\
& = & (\bar{t}^{\rm int}_L \overline{T}^{\rm int}_L)
\begin{pmatrix}
\frac{y_t}{\sqrt{2} }(h+v_H-i\phi^0) & \frac{\lambda_T}{\sqrt{2}}(h+v_H-i\phi^0)\\ 0 & y_T(s + v_S)
\end{pmatrix}
\begin{pmatrix}
t_R^{\rm int} \\ T_R^{\rm int} 
\end{pmatrix} \\
& \equiv & (\bar{t}_L \overline{T}_L)\,
\mathcal{U}_L (\mathcal{M} + \mathcal{H} + \mathcal{S})\,
\mathcal{U}_R^\dagger
\begin{pmatrix}
t_R \\ T_R 
\end{pmatrix}\\
& = & (\bar{t}_L \overline{T}_L)\, \mathcal{M}_{\rm diag} 
\begin{pmatrix}
t_R \\ T_R 
\end{pmatrix} +
(\bar{t}_L \overline{T}_L)\, \mathcal{U}_L (\mathcal{H} + \mathcal{S})\,
\mathcal{U}_R^\dagger
\begin{pmatrix}
t_R \\ T_R 
\end{pmatrix}, \label{eq:L_yuk} 
\end{eqnarray}
with $\mathcal{M}$ as in Eq.(\ref{eq:massmatrix}), and 
\begin{equation}
\mathcal{H} =
\frac{h-i\phi^0}{\sqrt{2}}\begin{pmatrix}
y_t  & \lambda_T \\
0 & 0 
\end{pmatrix}; \;
\mathcal{S} = s 
\begin{pmatrix}
0  & 0 \\
0 & y_T 
\end{pmatrix}.
\end{equation}
It is straightforward to work out the terms involving the scalar fields $h, s$: 
\begin{equation}
\mathcal{U}_L \mathcal{H}\mathcal{U}_R^\dagger = \frac{h-i\phi^0}{v_H}
\begin{pmatrix}
m_t c_L^2 & m_T s_L c_L \\ 
m_t s_L c_L & m_T s_L^2 
\end{pmatrix}\,,
\end{equation}
and 
\begin{equation}
\mathcal{U}_L \mathcal{S}\mathcal{U}_R^\dagger = \frac{s}{v_S}
\begin{pmatrix}
m_t s_L^2 & -m_T s_L c_L \\ 
-m_t s_L c_L & m_T c_L^2 
\end{pmatrix}.
\end{equation}
The terms involving the bottom quark are
\begin{equation}
\mathcal{L} \supset  -i \phi^- \bar{b}_L (y_t t_R^{\rm int} + \lambda_T T_R^{\rm int}) - iy_t \phi^+ \bar{t}_L^{\rm int}b_R+ \frac{y_b}{\sqrt2}(h+v_H+i\phi^0)\bar{b}_Lb_R.\\
\end{equation}

\subsection{Input Parameters}\label{sec:input}

The parameters in the potential (\ref{eq:potential}) are
\begin{equation}
\mu, \lambda, a_1, a_2, b_1, b_2, b_3, b_4.
\end{equation}
To reduce the number of free parameters we can use (\ref{eq:min1}) and (\ref{eq:min2}) to eliminate $\mu$ and $a_2$:
\begin{eqnarray}
\mu^2 & = & \lambda v_H^2 + \frac{a_1}{2}  v_S +  \frac{a_2}{2} v_S^2\,, \\
a_2 & = & -\frac{a_1}{2v_S} -  \frac{2}{v_H^2 v_S}\left(b_1 + b_2v_S + b_3 v_S^2 + b_4 v_S^3\right).
\end{eqnarray}
To proceed further, we note a simple relation between the mixing angle and the difference of the scalar masses:
\begin{equation}\label{eq:lamhs}
\sin(2 \theta) = \frac{2 M_{12}^2}{\sqrt{(M_{11}^2 - M_{22}^2)^2 + 4 M_{12}^4}} = \frac{(a_1 + 2 a_2 v_S) v_H}{m_2^2 - m_1^2}.
\end{equation}
Combining (\ref{eq:lamhs}) and (\ref{eq:min2}) we can eliminate $b_2$: 
\begin{equation}
b_2 = - \frac{v_H}{4v_S} \sin(2\theta) (m_2^2 - m_1^2)  - \frac{b_1}{ v_S} - b_3 v_S - b_4 v_S^2.
\end{equation}
 To eliminate $\lambda$ and $b_4$ we use 
\begin{eqnarray}
m_1^2 m_2^2 & = & \det(M) = M_{11}^2 M_{22}^2 - M_{12}^4\,, \\
m_1^2 + m_2^2 & = & M_{11}^2 + M_{22}^2.
\end{eqnarray}  
We find 
\begin{eqnarray}
m_1^2 \cos^2\theta + m_2^2 \sin^2\theta & \!\!= \!\! & M_{11}^2 \;=\; 2 \lambda v_H^2 \label{eq:lamh}\,, \\
m_1^2 \sin^2\theta + m_2^2 \cos^2\theta & \!\!= \!\! & M_{22}^2 \; = \; \frac{a_2}{2} v_H^2 + b_2 + 2 b_3 v_S + 3 b_4 v_S^2.
\end{eqnarray}
Note that there are two solutions for $\lambda$. We chose the solution (\ref{eq:lamh}) because it gives the SM relation $\lambda = m_1^2/(2 v_H)$ in the limit of no mixing, $\theta \to 0$. 

If we impose a $Z_2$ symmetry for the scalar field $S$, as in \cite{Batell:2012zw}, we have $a_1 = b_1 =  b_3 =0$, and the scalar part of the model is specified in terms of the masses $m_1, m_2$, the mixing angle $\theta$, and the vacuum expectation values $v_H, v_S$. Since this more restricted model is a good starting point for phenomenological studies, let us collect the relevant equations: 
\begin{eqnarray}
\mu^2 & = & \frac12 \left( m_1^2 \cos^2\theta + m_2^2 \sin^2\theta \right) + \frac14 \frac{v_S}{v_H}\sin(2 \theta) (m_2^2 - m_1^2)\,, \\
\lambda & = & \frac{1}{2 v_H^2}(m_1^2 \cos^2\theta + m_2^2 \sin^2\theta )\,, \\
a_2 & = & \frac{1}{2v_Hv_S}\sin(2 \theta) (m_2^2 - m_1^2) \,, \\
b_2 & = & -\frac12 (m_1^2 \sin^2\theta + m_2^2 \cos^2\theta) - \frac14 \frac{v_H}{v_S}\sin(2 \theta) (m_2^2 - m_1^2)\,, \\
b_4 & = & \frac{1}{2 v_S^2} \left(m_1^2 \sin^2\theta + m_2^2 \cos^2\theta\right).
\end{eqnarray}
Furthermore, the conditions for an absolute minimum of the potential are always fulfilled in this particular case: 
\begin{eqnarray}
\lambda = \frac{1}{2 v_H^2}(m_1^2 \cos^2\theta + m_2^2 \sin^2\theta ) & > & 0 \,, \\
\frac{a_2}{2}v_H^2 + b_2 + 2 b_3 v_S + 3 b_4 v_S^2 = m_1^2 \sin^2\theta + m_2^2 \cos^2\theta & > & 0 \,, \\
2 \lambda v_H^2 \left(\frac{a_2}{2} v_H^2 + b_2 + 2 b_3 v_S + 3 b_4 v_S^2 \right) - v_H^2 \left(\frac{a_1}{2} + a_2v_S\right)^2  = m_1^2 m_2^2 & > & 0.
\end{eqnarray}

The Yukawa sector, Eq.(\ref{eq:yuk}), is determined in terms of the couplings $y_t, y_T$ and $\lambda_T$.
Using
\begin{eqnarray}\label{eq:mixing_params}
{\rm Tr}\,[U_L\mathcal{M}\mathcal{M}^\dagger U_L^\dagger] & = & m_t^2 +  m_T^2\,, \\
{\rm Det}\,[U_L\mathcal{M}\mathcal{M}^\dagger U_L^\dagger] & = & m_t^2 m_T^2\,, \\
(U_L \mathcal{M}\mathcal{M}^\dagger U_L^\dagger)_{i,j, i\neq j} & = & 0\,,
\end{eqnarray}
we can express $y_t, y_T$ and $\lambda_T$ in terms of the top-quark masses, the mixing angle and the vacuum expectation values:
\begin{eqnarray}\label{eq:mixing_params1}
y_T^2 v_S^2 & = & \sin^2\theta_L m_t^2 + \cos^2\theta_L^2 m_T^2\,, \\
\frac12 y_t^2 y_T^2 v_H^2 v_S^2 & = & m_t^2 m_T^2\label{eq:mixing_params2}\,, \\
\frac12 y_t^2 v_H^2 + \frac12 \lambda_T^2 v_H^2 + y_T^2 v_S^2 & = & m_t^2 + m_T^2 \label{eq:mixing_params3}.
\end{eqnarray}

Combining eqs.(\ref{eq:mixing_params1}-\ref{eq:mixing_params3}) we find that the ratio of the Yukawa couplings $\lambda_t$ and $y_T$ is given by the mixing angle and the top masses: 
\begin{equation}
\frac{\lambda_T}{y_t} = \sin\theta_L \cos\theta_L \frac{m_T^2-m_t^2}{m_t m_T}.
\end{equation}
We would expect both couplings to be of $\mathcal{O}(1)$, and thus $m_t m_T \sim \sin\theta_L \cos\theta_L (m_T^2-m_t^2)$.

\bibliographystyle{JHEP}
\bibliography{simplified_higgs}
%\printbibliography

\end{document}